\documentclass[twocolumn,twocolappendix]{aastex631}

\usepackage{amsmath}

\usepackage{orcidlink}
\newcommand{\AuthorORCID}[2]{%
  \href{https://orcid.org/#2}{\mbox{#1~{\Large\orcidlink{#2}}\kern-0.5em}}%
  }

\usepackage{etoolbox}
\makeatletter
\patchcmd{\acknowledgments}
  {\begin{internallinenumbers}}
  {\ifnumlines\begin{internallinenumbers}\fi}
  {}
  {\GenericWarning{}{Could not patch acknowledgments start}}
\patchcmd{\endacknowledgments}
  {\end{internallinenumbers}}
  {\ifnumlines\end{internallinenumbers}\fi}
  {}
  {\GenericWarning{}{Could not patch acknowledgments end}}
\makeatother

\begin{document}

\title{High-Energy Neutrinos from Supernova Shock Breakouts in Circumstellar Media: Light Curves, Spectra, and Contribution to the Extragalactic Neutrino Background}

\shorttitle{High-energy Neutrinos from SN Shock Breakouts in CSM}
\shortauthors{Wasserman et al.}

\author{\AuthorORCID{Tal Wasserman}{0009-0005-7414-3965}}
\affiliation{Department of Particle Physics \& Astrophysics, Weizmann Institute of Science, Rehovot 76100, Israel}
\author{\AuthorORCID{Eli Waxman}{0000-0002-9038-5877}}
\affiliation{Department of Particle Physics \& Astrophysics, Weizmann Institute of Science, Rehovot 76100, Israel}
\author{\AuthorORCID{Kohta Murase}{0000-0002-5358-5642}}
\affiliation{Department of Physics; Department of Astronomy \& Astrophysics; Center for Multimessenger Astrophysics, Institute for Gravitation and the Cosmos, The Pennsylvania State University, University Park, PA 16802, USA}
\affiliation{Center for Gravitational Physics and Quantum Information, Yukawa Institute for Theoretical Physics, Kyoto University, Kyoto, Kyoto 606-8502, Japan}

\begin{abstract}
Enhanced mass loss from core-collapse supernova (SN) progenitors shortly before explosion appears to be common, creating a compact optically thick circumstellar medium (CSM) at $\sim10^{14}-10^{15}$ cm. We derive an analytic description of the light curves and spectra of high-energy neutrinos emitted by nonrelativistic SN shock breakouts through such CSM, as a function of shock velocity and CSM parameters, accounting for the evolution of the hydrodynamic structure and the electromagnetic (EM) spectrum as the shock transitions from being radiation-mediated to collisionless. This evolution determines the time-dependent neutrino production efficiency, the maximum proton/neutrino energy, and the pair-production optical depth. A significant fraction of the neutrino energy is typically emitted within a few days of explosion, during breakout and before the EM light curve peak, with $1-100$ TeV neutrinos carrying $\approx10\%$ of the energy of shock-accelerated protons. The escape of high-energy photons ($>1$~GeV) is suppressed by pair-production for compact CSM configurations. If enhanced mass losses are common, and assuming that shock-accelerated protons carry $\approx10\%$ of the collisionless shock energy, CSM SN breakouts may significantly contribute to the observed high-energy neutrino background, without overproducing a corresponding high-energy gamma-ray background. SNe producing $>1$ neutrino events in a $1\left(10\right){\rm km^2}$ detector are expected at a rate of $\sim0.05\left(1\right){\rm yr^{-1}}$. 
\end{abstract}

\section{Introduction}\label{sec:intro}

Early light curves of many core-collapse supernovae (SNe) are thought to be powered by the collision of the stellar ejecta with a compact, optically thick circumstellar medium (CSM) shell surrounding the progenitor star \citep[e.g.,][]{yaron_confined_2017,forster_delay_2018,morozova_measuring_2018,irani_early_2024,jacobson-galan_final_2024,hinds_inferring_2025}, likely ejected in the final years before explosion. Such a collision drives a radiation-mediated shock (RMS) in the CSM, which breaks out once the Thomson optical depth ahead of it becomes comparable to the optical depth of the shock width $\tau=c/v$, where $v$ is the shock velocity \citep{weaver_structure_1976}, and the photons escape as their diffusion time becomes shorter than the dynamical time, on a days timescale \citep[see][for reviews]{waxman_shock_2017,levinson_physics_2020}. 

During the shock breakout in the CSM, as the radiation is unable to efficiently accelerate the plasma, there is a transition of the shock structure from an RMS to a collisionless shock (CLS) \citep{katz_x-rays_2011,wasserman_optical_2025}. The CLS deposits most of the shock power directly in the thermal plasma (which cools quickly by radiating this energy), such that the RMS-to-CLS transition implies a significant increase in the temperature of the shock-heated plasma, from tens of eV to tens of keV \citep{katz_x-rays_2011}, shifting the spectrum's peak from the UV to the X-ray band on the breakout timescale of days, with a negligible suppression of the X-ray photons by the upstream plasma \citep{wasserman_optical_2025}. In addition, the CLS (unlike the RMS) is expected to produce a ``tail" of high-energy cosmic ray (CR) protons by diffusive shock acceleration \citep{krymskii_regular_1977,bell_acceleration_1978,blandford_particle_1978} to energies well above the $\sim10\,$keV temperature that characterizes the bulk of the shock-heated electrons and protons, which are expected in such dense environments to produce high-energy ($h\nu\gg m_ec^2$) gamma rays and multi-TeV neutrinos \citep{murase_new_2011,katz_x-rays_2011,kashiyama_high-energy_2013,murase_probing_2014,giacinti_collisionless_2015,zirakashvili_type_2016,petropoulou_point-source_2017,murase_new_2018,li_pev_2019,sarmah_high_2022,pitik_optically_2023,murase_interacting_2024,waxman_shock_2025,duan_probing_2026,gagliardini_ultraviolet_2026}.

\subsection{Preexplosion Mass Losses}
\label{sub-sec:mass-loss}

With the advent of new wide-field high-cadence sky surveys in recent years, early light curves are accessible for numerous events, contributing to the growing observational evidence that an enhanced mass loss during $\sim1\,$yr preceding the explosion is common in many, and perhaps most of the progenitors of core-collapse supernovae, creating a compact CSM shell at $\sim10^{14}-10^{15}\,$cm radii. The evidence for the prevalence of compact CSM around SNe progenitors includes ``flash spectroscopy" \citep[e.g.,][]{yaron_confined_2017}, analyses of emission properties of SNe samples with early optical+UV observations \citep[e.g.,][]{irani_early_2024}, systematic analyses of precursor bursts associated with preexplosion mass ejection events \citep[e.g.,][]{ofek_precursors_2014,strotjohann_bright_2021}, and detailed studies of the CSM structure of nearby SNe like SN 2023ixf \citep[see][for a more complete account and references of the observational evidence]{wasserman_optical_2025,waxman_shock_2025}.

Various suggestions have been made for the preexplosion mass loss origin, which is not yet well understood, including: pair instability pulsations \citep[e.g.,][]{rakavy_instabilities_1967,woosley_pulsational_2007}, binary interaction \citep[e.g.,][]{chevalier_common_2012,soker_explaining_2013}, radiation-driven instability \citep[e.g.,][]{suarez-madrigal_local_2013}, unstable late-stage nuclear burning \citep[e.g.,][]{smith_preparing_2014,woosley_remarkable_2015}, dissipation of internal gravity waves driven by core burning \citep[e.g.,][]{shiode_setting_2013,fuller_pre-supernova_2017,fuller_pre-supernova_2018}, and core magnetic activity \citep[][]{cohen_pre-supernova_2024}. Some of these mechanisms result in an inflated, gravitationally bound stellar envelope, rather than a mass ejection. This distinction would not be important for the EM and neutrino emission as long as the density structure is similar. Intense mass losses or inflation episodes challenge the canonical picture \citep[e.g.,][]{langer_presupernova_2012} of a rapidly evolving core surrounded by a nearly time-independent envelope. Information on the progenitor star's structure and mass loss history near the explosion is therefore highly instructive for the study of SN explosion mechanisms, which remain poorly understood despite many years of research.

\subsection{The Neutrino Background and CSM SN Breakouts}
\label{sub-sec:Nu-bgnd}

The origin of the astrophysical high-energy, $\gtrsim10\,$TeV, neutrino background \citep{icecube_collaboration_evidence_2013,IceCubeLargeENB2015,fusco_study_2019,IceCubeLargeENB2020,IceCubeLargeENB2026} remains one of the outstanding puzzles in the field, and its isotropy suggests that extragalactic sources dominate it. The $\gtrsim100$\,TeV neutrino intensity, is comparable to the Waxman-Bahcall (WB) bound \citep{waxman_high_1998}, suggesting a possible connection to ultrahigh-energy CR sources, though no direct experimental evidence is yet at hand. The reported 10--30\,TeV neutrino intensity, $\varepsilon_\nu^2 dI/d\varepsilon_\nu\sim10^{-7}$\,GeV\,cm$^{-2}$\,s$^{-1}$\,sr$^{-1}$ is a few times larger than the WB bound, and is comparable to the $\sim100$ GeV gamma-ray background \citep{ackermann_spectrum_2015}. The similarity to the $100\,$GeV gamma-ray background could, in principle, suggest a common source origin, as pionic neutrino production is accompanied by gamma-ray production at a similar rate, leading (after electromagnetic cascades from interactions with the IR background) to a comparable $100$ GeV gamma-ray intensity. However, it was found that the measured 10--30\,TeV neutrino intensity is actually in some tension with the $100\,$GeV gamma-ray intensity \citep{murase_hidden_2016}. 
This is mainly because the $100\,$GeV gamma-ray background is already largely accounted for by blazars' emission \citep{ackermann_resolving_2016}, which were not found to dominate the neutrino background \citep{aartsen_contribution_2017}. 
This tension has driven interest in sources of the 10--30\,TeV neutrinos that are \emph{opaque} to $100$ GeV gamma rays, including ``choked-jets" in exploding massive stars \citep{meszaros_tev_2001,murase_tev-pev_2013,denton_bright_2018,zegarelli_towards_2024}, super/hyper-Eddington accretion onto compact objects \citep{sridhar_high-energy_2024}, and the vicinity of active galactic nuclei \citep{stecker_high-energy_1991,kalashev_neutrinos_2015,murase_hidden_2020}.

SN shocks driven into the interstellar medium or stellar winds surrounding massive stars have long been suggested as sources of Galactic CRs \citep{blandford_particle_1987}, but the high-energy neutrino emission due to energy losses by these CRs (which extends over hundreds of days), is too low to account for the neutrino background, and it is accompanied by a similar high-energy gamma-ray emission since the pair-production optical depth is small. This is because a significant fraction of the SN kinetic energy is converted into CRs around the Sedov time, at which the density is sufficiently low.

However, the prevalence of dense CSM in common core-collapse SNe implies that high-energy neutrino sources may be ubiquitous. Current high-energy neutrino detectors may detect the next Galactic SN \citep{murase_new_2018}, but are not sensitive enough for the detection of the high-energy neutrino signal from the closest extragalactic CSM SN breakouts, including SN 2023ixf and 2024ggi at $\sim7\,$Mpc \citep[Section \ref{sec:discussion} and][]{murase_interacting_2024,sarmah_new_2024,waxman_shock_2025,cosentino_high-energy_2025,kimura_high-energy_2025,buccheri_high-energy_2026}. While possible associations of high-energy neutrino events with SNe have been recently suggested \citep{lu_sn2023syz_2025,stein_sn_2025,ji_high-energy_2026,garrappa_type_2026,Sawada_2026,lu_high-energy_2026}, the statistical significance and physical interpretation of these associations remain uncertain.

In \citet{waxman_shock_2025} we showed that if enhanced mass losses from the progenitors of normal core-collapse SNe, that occur at a rate of $\approx10^{-4}\,{\rm Mpc^{-3}\,yr^{-1}}$, and have outer ejecta shock velocities of $v\sim10^9\,{\rm cm\,s^{-1}}$, are indeed common, then the breakouts of the SN shocks through the compact CSM shells produce a neutrino flux that may account for a significant fraction of the observed $\gtrsim$10 TeV neutrino background,
\begin{equation}
    \begin{split}
            \varepsilon_\nu^2\frac{dI_\nu}{d\varepsilon_\nu}\sim&10^{-7}\frac{\rm GeV}{\rm cm^2\,s\,sr} \left(\frac{M_{\rm e}}{10^{-1}\,M_\odot}\right)\left(\frac{v}{10^9\,\rm cm\,s^{-1}}\right)^2\\
            &\times\left(\frac{\epsilon_{\rm CR}}{10^{-1}}\right)\left(\frac{R_{\rm SN}}{10^{-4}\,{\rm Mpc}^{-3}\,{\rm yr}^{-1}}\right),
    \end{split}
\end{equation}
where $M_{\rm e}$ is the dense CSM mass, $\epsilon_{\rm CR}$ is the CR energy fraction, and $R_{\rm SN}$ is the local SN rate \citep[see also findings by][]{sarmah_high_2022}. For compact CSM distributions, the neutrinos are produced over the few-day timescale, coincident with the bright UV (possibly followed by X-ray) breakout emission. Normal type II SN explosions that occur at a rate of $\approx10^{-4}\,{\rm Mpc^{-3}\,yr^{-1}}$, would contribute $\approx30\%\left(M_{\rm e}/0.03\,M_\odot\right)$ of the observed background, and type IIn SNe, that occur at a rate of $\approx10^{-5}\,{\rm Mpc^{-3}\,yr^{-1}}$ and show prevalent ejection of a fraction of $M_\odot$ preceding the explosion, may produce another $\approx 30\%(M_{\rm e}/0.3\,M_\odot)$ of the background.\footnote{Some works \citep[e.g.,][]{zirakashvili_type_2016,petropoulou_point-source_2017,pitik_optically_2023,salmaso_diversity_2025} focused on spectacular, yet very rare SNe interacting with $\sim10\,M_\odot$ CSM mass \citep[e.g., SN 2010jl][]{ofek_sn_2014} (which significantly decelerates the ejecta). We note that the rate of these events is very low compared to that of the common Type IIn SNe ($\sim 10^{-5}\,{\rm Mpc}^{-3}\,{\rm yr}^{-1}$), and they are unlikely, therefore, to contribute significantly to the neutrino background.} Stacking analyses have placed upper limits on the energy emitted in high-energy neutrinos following SNe \citep{abbasi_constraining_2023}, which are still large enough to leave room for SNe to dominate the background signal, and anyway they are not directly applicable to the scenario that we consider, as they were obtained for neutrino emission over long time windows (100 days and longer), starting at the time of the first optical SN detection, which is typically later than the few days' timescale over which most of the neutrino emission is predicted to occur.

The small contribution of SNe with dense CSM to the neutrino background argued by \citet{duan_probing_2026} appears to be driven mainly by their assumed CSM population, which they inferred from optical samples. As most of the early radiation from CSM interaction is emitted in the UV, optical light curve analyses do not constrain the progenitor system properties well \citep{wasserman_supernovae_2026} and are expected to be highly degenerate with the optical radiation arising from the cooling stellar envelope beneath the CSM. The optically-inferred small CSM masses are in tension with analyses of early SN light curves that include the UV bands \citep[e.g.,][]{irani_early_2024, jacobson-galan_final_2024}.

\subsection{Current Work}
\label{sub-sec:This-paper}

Here, we extend the analysis of previous studies of neutrino emission in CSM SN breakouts by accounting for the hydrodynamic evolution of the shock structure and for the evolution of the radiation field through the RMS-to-CLS transition (we leave the determination of the luminosity and spectrum of the escaping high-energy, $> m_ec^2$, gamma rays, as well as the distortions of the low-energy photon spectrum by the emission of CR electrons, to future work). The hydrodynamic evolution determines the onset and efficiency of proton acceleration, the efficiency of neutrino production, and the temporal structure of the neutrino signal. It is particularly important for compact CSM configurations, which appear to be common and may apply to SN 2023ixf (see Section~\ref{sec:discussion}), where the dense CSM extends only modestly beyond the breakout radius and the RMS-to-CLS transition occurs close to the dense CSM edge. The evolution of the radiation field, which departs from thermal equilibrium and shifts from UV to X-ray energies during the RMS-to-CLS transition, plays a central role in shaping the high-energy emission: it limits the maximum proton, and hence neutrino, energy through $p\gamma$ pion-production, and it determines the suppression of gamma-ray escape through $\gamma\gamma$ pair-production. Earlier works using simplified treatments of the photon density and energy distribution (typically assuming thermal-equilibrium), do not account properly for these important effects.

The current analysis is based on our recent solutions to the problem of shock breakouts within a CSM ``wind" density profile, $\rho\propto r^{-2}$, describing the evolution of the plasma and radiation field structure, including the transition from RMS to CLS, and the transition from UV to X-ray emission \citep{wasserman_optical_2025}. A theoretical description of the evolution of the EM spectrum during breakout was previously unavailable, as it was challenged by the formation of the CLS (which was not included in most works), by the nonsteady nature of the problem (with shock structure evolving over a dynamical time, rendering steady-state solutions inapplicable and greatly complicating analytic analyses), by the deviation from a thermal equilibrium, and by the importance of inelastic Compton scattering in determining the (quasithermal) electron temperature and the optical-X-ray spectrum. Our numerical code solves the 1D spherically symmetric multi-group radiation-hydrodynamics equations using the diffusion approximation for radiation transport (valid for the $v/c\ll 1$ shocks considered), including bremsstrahlung emission/absorption and inelastic Compton scattering, described by the Kompaneets equation. The formation and evolution of the CLS is captured by an artificial viscosity term (a valid approximation as the CLS width, of the order of the plasma skin depth, is very small compared to all other length scales of the problem), and a ``cooling limiter" is introduced to capture the correct shock-heated electron temperature and post-shock cooling profile with acceptable grid resolution. The numerical results were confirmed against analytic expressions available in limiting parameter regimes. This framework enables us to solve the CSM breakout problem self-consistently, addressing the challenges described above.

Building on our full radiation-hydrodynamics solutions, we quantitatively analyze the proton acceleration and the high-energy neutrino emission from CSM shock breakouts. We derive scaling relations for the neutrino light curves, timescales, and energies, presenting the dependence of the emission properties on the CSM parameters, and analyze the possible contribution to the neutrino background. A description of the relation between the light curves and spectra of the electromagnetic (EM) and neutrino emissions is crucial for both inference of progenitor parameters and for dedicated neutrino stacking searches (see Sections \ref{sec:nu_em_relation} and \ref{sec:discussion}).

The key approximations we use are as follows.
\\ \noindent [1] \textit{Nonthermal particles and magnetic fields.} A complete first-principles understanding of the CLS structure, particularly of the fractions $\epsilon_\mathrm{CR}$, $\epsilon_\mathrm{B}$ of post-shock internal energy carried by CRs and magnetic fields, and of the spectral index $s$ of the CR momentum distribution, is not yet available \citep[e.g.,][for reviews]{gupta_ab-initio_2023,sironi_relativistic_2015}. We assume that the energy fractions are constant in time and space and treat them as model parameters, and adopt a flat spectrum $s=2$. While observations of SN remnants suggest a somewhat steeper index of $s=2.2-2.3$ \citep[e.g.,][]{ackermann_detection_2013}, our results are not very sensitive to such a change in the proton spectrum, as discussed in Sections \ref{sec:nu_emission}-\ref{sec:nu_background}.
\\ \noindent [2] \textit{The impact of nonthermal particles on shock dynamics and radiation field.} The CRs accelerated by the CLS are not expected to significantly modify the shock dynamics or the optical-to-X-ray radiation field, so in the calculation of \citet{wasserman_optical_2025}, the dynamics of the flow and radiation field are solved neglecting the CRs. While CRs may affect the dynamics through escape of neutrinos, through generation of gamma rays for which the scattering cross-section differs from Thomson's, and through modification of the optical depth by pair-production, these effects are small: neutrinos carry only $0.05\,\epsilon_{\rm CR,-1}$ of the energy, and the dynamical effect of gamma rays is small since only a fraction of the CR energy is radiated at $h\nu\gg m_ec^2$ and since the pair-production optical depth is significant only for $h\nu\gg m_ec^2$ photons, limiting their propagation (the resulting pairs lose their energy rapidly radiating lower energy photons), and the contribution of pairs to the optical depth is small for $v/c<0.1$ \citep{katz_fast_2010}. Finally, the electron downstream temperature, determined by the balance of radiative cooling to proton heating, is not significantly modified since the energy density of $h\nu<m_ec^2$ photons is not significantly modified. 
\\ \noindent [3] {\it CSM profile.} While observations do not yet provide stringent constraints on the CSM structure or the ejection mechanisms, the results of numerical simulations of mass ejection following energy deposition in stars yield density distributions which are not very different from $\rho\propto r^{-2}$ profile \citep[][and references therein]{wasserman_optical_2025}. In addition, most of the dynamical key features are insensitive to the exact details of the density profile, and we therefore analyze a simple CSM setting of $\rho\propto r^{-2}$ with a finite extent, which enables a clear analytic understanding of the EM and neutrino emission and their CSM mass and radius dependence, which can be easily generalized for a wide range of density profiles (Section \ref{sec:formulation} and \citet{wasserman_supernovae_2026}).
\\ \noindent [4] {\it Shock velocity.} The shocked expanding stellar envelope acts as a “piston” driving a shock through the CSM. For small CSM mass $M_{\rm e}\ll1\,M_\odot$, the shock is driven by the fast outer shell of the envelope. For hydrostatic stellar envelopes, this shell is typically accelerated above $10^9\,\rm cm\,s^{-1}$ \citep{matzner_expulsion_1999}, with a shallow dependence on mass coordinate $v\propto m^{-0.11}$, such that for $\rho\propto r^{-2}$, the outer envelope/shock velocity decreases slowly $v\propto t^{-0.11}$ (and the density of the expanding envelope shell is much larger than the density of the CSM, resulting in a weak reverse shock driven back into the envelope, see explanation in \cite{wasserman_optical_2025}). We therefore approximate the shock velocity $v=10^9\,{\rm cm\,s^{-1}}\,v_9$ to be constant in time throughout this work, although our result can be easily generalized to include any velocity as a function of time (in the velocity validity range, see Section \ref{sec:formulation}).
\\ \noindent [5] {\it Emission from the expanding stellar envelope.} For sufficiently extended CSM distribution, the luminosity from the expanding stellar envelope beneath the CSM can overwhelm the luminosity from CSM interactions and dominate the radiation energy density. This will be the case for CSM of mass $<10^{-2}-10^{-1}\,M_\odot$ at radii $>10^{16}-10^{17}\,{\rm cm}$, and interaction timescales of months to years. As observations suggest that the common CSM configuration is more compact, we leave the discussion of this part of the parameter range to future work.

The paper is organized as follows. In Section~\ref{sec:formulation}, we present the CSM breakout setting, explain how we use the hydro-radiation solution of \citet{wasserman_optical_2025}, and analyze the corresponding proton acceleration and neutrino production. Section~\ref{sec:nu_emission} presents the resulting neutrino light curves and spectra and their dependence on parameters. In Section~\ref{sec:nu_background}, we estimate the contribution to the extragalactic neutrino background for different system parameters. 
In Section~\ref{sec:g_supp}, we calculate the $\gamma\gamma$ optical depth and determine under which conditions high-energy gamma rays can escape. Section~\ref{sec:nu_em_relation} discusses the relation between neutrino and EM emission.  We summarize and discuss our results in Section~\ref{sec:discussion}.

\section{CSM Breakouts, Proton Acceleration and Neutrino Production}\label{sec:formulation}

In this section, we present the CSM breakout setting, explain how we use the radiation-hydrodynamics solution of \citet{wasserman_optical_2025} (Sections \ref{subsec:sbo}-\ref{subsec:EM_rad}), and analyze the corresponding proton acceleration (Section \ref{subsec:proton_acceleration}) and neutrino production (Section \ref{subsec:nu_prod}). For the hydrodynamics description, we mostly employ the derived analytic results (that were numerically validated), while for the spectral energy distribution of the radiation, we use tabulated numerical results (since analytic descriptions of the spectrum are available only in limiting regimes of the physical parameter space). We also account for the transition of the shock from the radiative to the adiabatic regime. Our treatment is expected to be accurate at radii not too close to the CSM edge $R_{\rm e}$, and for $0.5<v_9<2$ (see below).

Figure \ref{fig:profiles} shows the evolution of several quantities derived below, evaluated at the shock location as it propagates through the CSM. A relatively small mass, extended CSM configuration is presented to demonstrate the general profile structure of the different quantities.

\begin{figure}
    \centering
    \includegraphics[width=8.5cm]{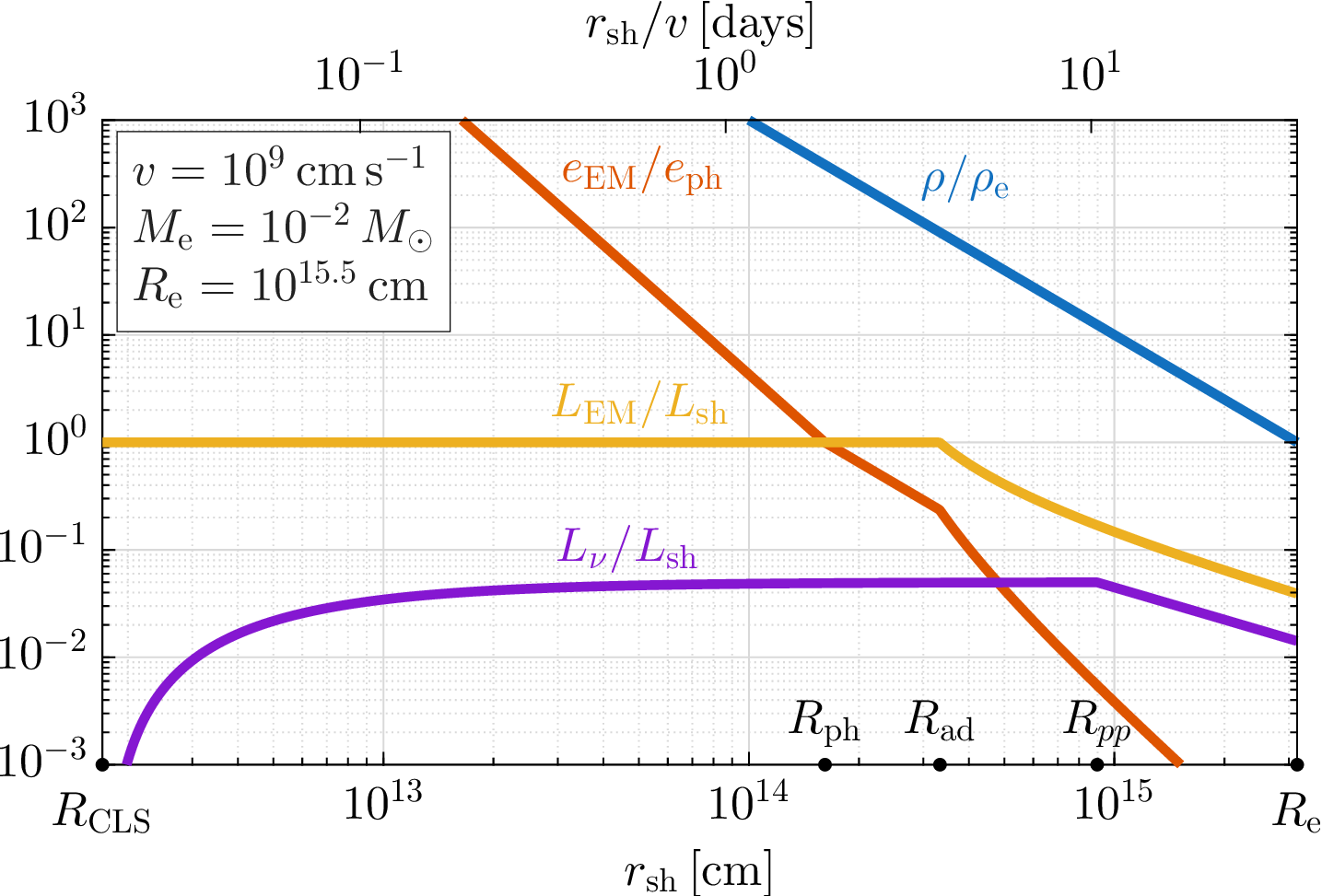}
    \caption{The evolution of the CSM density $\rho$, Equation (\ref{eq:density}), EM radiation energy density $e_{\rm EM}$ and luminosity $L_{\rm EM}$, Equation (\ref{eq:e_EM}), and neutrino luminosity $L_\nu$, Equation (\ref{eq:L_nu}), at the shock location. Characteristic radii are marked on the $x$ axis: The CLS onset radius $R_{\rm CLS}$, Equation (\ref{eq:R_CLS}), the photospheric radius $R_{\rm ph}$, Equation (\ref{eq:R_bo}), the adiabatic radius $R_{\rm ad}$, Equation (\ref{eq:R_ad}), and the $pp$-inefficiency radius $R_{pp}$, Equation (\ref{eq:R_pp}).  The energy density is normalized to its photospheric value, and the luminosities are normalized to the shock power $L_{\rm sh}$, Equation (\ref{eq:L_sh}). The upper $x$ axis shows the corresponding time $t=r_{\rm sh}/v$. We use $\kappa_{\rm T}=0.34\,{\rm cm^2\,g^{-1}}$, $\epsilon_{\rm CR}=10^{-1}$ and $\epsilon_B=10^{-2}$.}
    \label{fig:profiles}
\end{figure}

We consider a spherically symmetric CSM of total mass $M_{\rm e}$ and constant opacity $\kappa_{\rm T}$ extending up to radius $R_{\rm e}$, with a density profile (see Section \ref{sec:intro})
\begin{equation}
\label{eq:density}
\begin{split}
    &\rho(r<R_{\rm e})=\rho_{\rm e}\left(\frac{r}{R_{\rm e}}\right)^{-2}=\frac{M_{\rm e}}{4\pi R_{\rm e}^3}\left(\frac{r}{R_{\rm e}}\right)^{-2}=\frac{\dot{M}_{\rm ou}}{4\pi r^2 v_{\rm ou}},\\
    &\rho_{\rm e}=1.6\times10^{-14}\,{\rm g\,cm^{-3}}\left(\frac{M_{\rm e}}{10^{-1}\,M_\odot}\right)\left(\frac{R_{\rm e}}{10^{15}\,\rm cm}\right)^{-3},
\end{split} 
\end{equation}
where the inner boundary radius of the CSM is assumed to be small. Such a profile could result from an outburst with a constant mass loss rate $\dot{M}_{\rm ou}$ at constant velocity $v_{\rm ou}$, initiated at a time $t_{\rm pr}=3.2\,{\rm yr}\left(R_{\rm e}/10^{15}\,\rm cm\right)\left(v_{\rm ou}/10^7\,{\rm cm\,s^{-1}}\right)^{-1}$ preceding the explosion, such that $M_{\rm e}=10^{-1\,}M_\odot\left(\dot{M}_{\rm ou}/10^{-1}\,M_\odot\,{\rm yr}^{-1}\right)\left(t_{\rm pr}/1\,{\rm yr}\right)$. 
The shock then sweeps the CSM on a timescale $R_{\rm e}/v=12\,{\rm d}\left(R_{\rm e}/10^{15}\,\rm cm\right)v_9^{-1}$ (justifiably neglecting the unshocked CSM velocity).

\subsection{Shock Breakout Dynamics}\label{subsec:sbo}

The power of a strong shock with a constant velocity in a $\rho\propto r^{-2}$ profile is time independent\footnote{We neglect the evolution of the adiabatic index from $4/3$ of the RMS to $5/3$ of the CLS (and use $5/3$ throughout), as it has a small effect on the results. For the luminosity, the additional $4\gamma/(\gamma+1)^2$ factor is anyway very close to unity and is omitted.}
\begin{equation}
\label{eq:L_sh}
\begin{split}
    &L_{\rm sh}(r)=4\pi r^2\times\frac{1}{2}\rho v^3=\frac{\frac{1}{2}M_{\rm e}v^2}{R_{\rm e}/v}\\
    &=10^{44}\,{\rm erg\,s^{-1}}\,v_{9}^3\left(\frac{M_{\rm e}}{10^{-1}\,M_\odot}\right)\left(\frac{R_{\rm e}}{10^{15}\,{\rm cm}}\right)^{-1}.
\end{split}
\end{equation}

The characteristic CSM optical depth is defined as
\begin{equation}
\label{eq:tau_e}
       \tau_{\rm e}\equiv\frac{\kappa_{\rm T} M_{\rm e}}{4\pi R_{\rm e}^2}=5.4\,\kappa_{0.34}\left(\frac{M_{\rm e}}{10^{-1}\,M_\odot}\right)\left(\frac{R_{\rm e}}{10^{15}\,\rm cm}\right)^{-2},
\end{equation}
with $\kappa_{0.34}=\kappa_{\rm T}/\left(0.34\,{\rm cm^2\,g^{-1}}\right)$. Note that $\tau_{\rm e}$ is not the total optical depth of the material, but rather the optical depth at radii $r\sim R_{\rm e}$ (for a finite $\rho\sim r^{-2}$ profile, it is the optical depth at $r=R_{\rm e}/2$). It is a measure of the ``compactness" of the CSM.

The breakout radius $R_{\rm bo}$, defined as the radius where the optical depth ahead of the shock drops to $c/v$, and the photosphere radius $R_{\rm ph}$ (where the optical depth drops to unity) are
\begin{equation}
\label{eq:R_bo}
    \begin{split}
        R_{\rm bo}&=\left(1+\frac{c/v}{\tau_{\rm e}}\right)^{-1}R_{\rm e},\\
        R_{\rm ph}&=\left(1+\frac{1}{\tau_{\rm e}}\right)^{-1}R_{\rm e}.
    \end{split}
\end{equation}
The limits $c/v\ll\tau_{\rm e}$ and $\tau_{\rm e}\ll c/v$ separate between ``edge" and ``interior" breakout regimes \citep{chevalier_shock_2011,khatami_landscape_2024,wasserman_supernovae_2026}. For an infinite CSM extent, with $R_{\rm e}\to\infty$ but with a fixed density normalization $M_{\rm e}/R_{\rm e}^3=\rm const.$, the breakout radius converges to $R_{\rm bo}^\infty$,
\begin{equation}
\label{eq:R_bo^inf}
\begin{split}
    &R_{\rm bo}<R_{\rm bo}^\infty=\frac{\tau_{\rm e}}{c/v}R_{\rm e}\\
    &=1.8\times10^{14}\,{\rm cm}\,\kappa_{0.34}\,v_9\left(\frac{M_{\rm e}}{10^{-1}\,M_\odot}\right)\left(\frac{R_{\rm e}}{10^{15}\,\rm cm}\right)^{-1},
\end{split}
\end{equation}
where for finite radial extent CSM with $M_{\rm e}=10^{-1}\,M_\odot$ and $R_{\rm e}=10^{15}\,{\rm cm}$, $R_{\rm bo}=1.5\times10^{14}\,{\rm cm}$. The quantity $R_{\rm bo}^\infty$ is meaningful also in the case of finite radial extent CSM, as it corresponds to the density normalization, $D\equiv \rho\left(r\right) r^2= R_{\rm bo}^\infty \left(c/v\right)/\kappa_{\rm T}=\dot{M}_{\rm ou}/\left(4\pi v_{\rm ou}\right)$.

During the breakout, as the shock propagates outward, it transitions from RMS to CLS. At small shock radii, upstream-diffusing radiation accelerates the plasma to near the post-shock velocity, whereas at larger shock radii, comparable to the breakout radius, the velocity evolution within the shock transition region can be divided into two parts: an outer, smooth-profile RMS acceleration followed by an abrupt CLS acceleration. Analytically, the start of the RMS-to-CLS transition or the ``CLS onset" radius is expected when the shock reaches $R_{\rm CLS}=R_{\rm bo}^\infty/2$, and was found numerically to be \citep{katz_x-rays_2011,wasserman_optical_2025}
\begin{equation}
\label{eq:R_CLS}
    R_{\rm CLS}\approx0.3\,R_{\rm bo}^\infty,
\end{equation}
which is somewhat smaller than $R_{\rm bo}^\infty/2$, as expected. The condition for a CLS phase to exist before the shock reaches $R_{\rm e}$ is thus $\tau_{\rm e}/\left(c/v\right)<3.3$, or, equivalently, $R_{\rm e}>2.3\times10^{14}\,{\rm cm}\,\kappa_{0.34}^{1/2}v_9^{1/2}\left(M_{\rm e}\right/10^{-1}\,M_\odot)^{1/2}$, or $R_{\rm e}/R_{\rm bo}>1.3$.

After the shock propagates beyond $R_{\rm CLS}$, the velocity jump due to the CLS evolves as \citep{wasserman_optical_2025}
\begin{equation}\label{eq:v_CLS}
    \tilde{v}_{\rm CLS}\left(r_{\rm sh}\right)\equiv v_{\rm CLS}\left(r_{\rm sh}\right)/v=1-\frac{R_{\rm CLS}}{r_{\rm sh}},
\end{equation}
and the acceleration quickly becomes CLS-dominated.

\subsection{Electromagnetic Radiation}\label{subsec:EM_rad}

The photon diffusion time from $R_{\rm bo}$ to $R_{\rm ph}$ \citep{ginzburg_superluminous_2012,wasserman_supernovae_2026}
\begin{equation}\label{eq:t_diff}
    \Delta t_{\rm diff}=2\left(\ln\frac{\tau_{\rm e}+c/v}{\tau_{\rm e}+1}+\frac{\tau_{\rm e}+1}{\tau_{\rm e}+c/v}-1\right)\frac{R_{\rm bo}^\infty}{v},
\end{equation}
reproduces the known limits for the breakout luminosity rise time of $R_{\rm bo}/v$ in the interior breakout limit and $\Delta R/v\equiv(R_{\rm ph}-R_{\rm bo})/v$ in the edge breakout limit.\footnote{For edge breakouts, the luminosity rise time may be smeared in general if the width over which the density decreases significantly beyond $R_{\rm e}$ is large, or by light travel time if $\tau_{\rm e}>\left(c/v\right)^2$  \citep{wasserman_supernovae_2026}. As we are interested in the parameter range where the CLS phase exists ($\tau_{\rm e}/\left(c/v\right)<3.3$), the second effect is not relevant here.}

The observed breakout peak of the EM emission (relative to the explosion time\footnote{Assuming that the time at which the shock reaches the inner boundary of the CSM is much shorter than $R_{\rm bo}/v$.}) is thus obtained at
\begin{equation}
    \label{eq:t_EM}
    t_{\rm EM}=R_{\rm bo}/v+\Delta t_{\rm diff}-R_{\rm ph}/c.
\end{equation}

The EM luminosity, and the EM energy density in the vicinity of the shock are given by\footnote{As both $\epsilon_{\rm CR}$ and $\epsilon_{\rm B}$ are small, we approximate the energy fraction carried by the thermal plasma as unity.}
\begin{equation} \label{eq:e_EM}
    \begin{split}
        L_{\rm EM}\left(r_{\rm sh}\right)&=L_{\rm sh}\left(r_{\rm sh}\right)\frac{1}{1+\frac{t_{\rm cool}}{t_{\rm dyn}}\left(r_{\rm sh}\right)},\\
        e_{\rm EM}\left(r_{\rm sh}\right)&=\frac{1}{2}\rho\left(r_{\rm sh}\right)\frac{v^3}{c}\frac{\max\left[1,\tau\left(r_{\rm sh}\right)\right]}{1+\frac{t_{\rm cool}}{t_{\rm dyn}}\left(r_{\rm sh}\right)}.
    \end{split}
\end{equation}
This expression accounts for both the enhancement of the energy density due to diffusion below the photosphere\footnote{An easy way to see this is that for uniform luminosity in space, that is obtained after breakout, the diffusion equation $\left(\partial_r e\right)/\rho\sim L/r^2$ gives $e\propto r^{-3}$ instead of $r^{-2}$ \citep[see][for a derivation and numerical validation of the evolution of the radiation energy density profile obtained in the downstream and upstream regions]{wasserman_optical_2025}. Note that this is an approximation in the intermediate, $\tau\sim1$ regime, but it is exact in the $\tau\gg1$ and $\tau\ll1$ limits (and for a general power law density $\rho\sim r^{-w}$ the enhancement for $\tau\gg1$ is $3\left(w-1\right)/\left(w+1\right)\tau$). Before the breakout, of course, this enhancement does not exist, such that at the scale of $\tau=c/v$ we get $e\sim\rho v^2$, which is the appropriate relation when the energy is trapped behind the shock, for $\tau\gtrsim c/v$.}, where $\tau>1$, and the suppression of the energy density at sufficiently large radii (which occurs for sufficiently large $R_{\rm e}$, see below) where the shock becomes adiabatic rather than radiative, i.e. when the shock-heated plasma radiative-cooling time, $t_{\rm cool}$, becomes long compared with the dynamical time, $t_{\rm dyn}$. A useful approximation that we use for this suppression is $\left(1+t_{\rm cool}/t_{\rm dyn}\right)^{-1}\approx\min\left(1,t_{\rm dyn}/t_{\rm cool}\right)$.

The plasma radiative-cooling time is defined as
\begin{equation}
\label{eq:t_cool}
t_\mathrm{cool}=\frac{e_\mathrm{pl}}{\dot{e}_\mathrm{Comp}+\dot{e}_\mathrm{brem}}\propto\frac{1}{1+Q_{\rm brem}},
\end{equation}
where $e_\mathrm{pl}$ is the shock-heated plasma energy density, $\dot{e}_\mathrm{brem}$ and $\dot{e}_\mathrm{Comp}$ are the bremsstrahlung and  Compton emissivities, and $Q_{\rm brem}\left(r_{\rm sh}\right)$ is their ratio, which was shown to play an important role in determining the radiation energy spectrum evolution \citep{wasserman_optical_2025}. At the adiabatic phase, defined as $t_{\rm dyn}<t_{\rm cool}$, the radiation energy density itself depends on the cooling time, and the cooling time depends on the radiation energy density (through the Compton emissivity), so there is an implicit relation that could be solved analytically for the cooling efficiency\footnote{The limit $\tilde{v}_{\rm CLS}\rightarrow1$ is used here since $R_{\rm ad}\gg R_{\rm CLS}$, and it is assumed that the shock-heated plasma temperature is much higher than the radiation temperature \citep[see][]{wasserman_optical_2025}.}
\begin{equation}
    \label{eq:t_dyn/t_cool}
    \begin{split}
    &\frac{t_{\rm dyn}}{t_{\rm cool}}\left(r_{\rm sh}>R_{\rm ad}\right)=\frac{\sqrt{\frac{3}{4\pi^3}}\frac{\left(\gamma+1\right)^4}{\gamma\left(\gamma-1\right)}\alpha_e\left(\frac{m_e}{m_p}\right)^{3/2}\left(\frac{c}{v}\right)^4}{\left(r_{\rm sh}/R_{\rm bo}^\infty\right)/\left[\frac{8}{3}\frac{m_p}{m_e}\frac{\gamma}{\left(\gamma+1\right)^2}\frac{v}{c}\right]-1}\\
    &=\frac{0.53\,v_9^{-4}}{\left(r_{\rm sh}/R_{\rm bo}^\infty\right)/\left(38\,v_9\right)-1}\overset{r_{\rm sh}\gg R_{\rm ad}}{\rightarrow} 20\left(\frac{r_{\rm sh}}{R_{\rm bo}^\infty}\right)^{-1}v_9^{-3},
    \end{split}
\end{equation}
where
\begin{equation}
    \label{eq:R_ad}
    R_{\rm ad}=38\,v_9\left(1+0.53\,v_9^{-4}\right)R_{\rm bo}^\infty,
\end{equation}
$\gamma=5/3$ being the adiabatic index.
The adiabatic region exists for sufficiently extended CSM of $R_{\rm e}>2.6\times 10^{15}\,{\rm cm}\,v_9\left(1+0.53\,v_9^{-4}\right)^{1/2}\left(M_{\rm e}/10^{-1}\,M_\odot\right)^{1/2}$, or $\tau_{\rm e}/\left(c/v\right)<\left(38\,v_9\right)^{-1}\left(1+0.53\,v_9^{-4}\right)^{-1}$. Note that $R_{\rm ad}$ is always beyond $R_{\rm ph}$ (Equation \ref{eq:R_bo}), so that, in general, the radiation energy density, Equation (\ref{eq:e_EM}), is characterized by three distinct regimes, see Figure \ref{fig:profiles}. The radiation energy density at the photosphere is $e_{\rm ph}=230\,{\rm erg\,cm^{-3}}\left(1+\tau_{\rm e}^{-1}\right)^2v_9^3\left(\rho_{\rm e}/10^{-14}\,{\rm g\,cm^{-3}}\right)$.

The time-dependent spectral energy distributions of the shock radiation are taken from the numerical tabulated results of the infinite wind-CSM solutions of \citet{wasserman_optical_2025}, with a set of $\{v,M_{\rm e},R_{\rm e}\}$ corresponding to $\{v,R_{\rm bo}^\infty\}$ through Equation (\ref{eq:R_bo^inf}). These solutions are expected to provide a good approximation also for the truncated CSM cases, at radii not too close to the edge, $r<R_{\rm e}$. In addition, while the numerical solutions obtained in the diffusion approximation are available only up to the photosphere, the spectrum is already well converged with respect to shock radius, so using the photospheric spectrum at larger radii provides a good approximation. The numerical spectra solutions are given for the velocity validity range $0.5<v_9<2$, such that relativistic effects of higher velocities, and UV/X-ray suppression effects of lower velocities, can be safely neglected, as shown in \cite{wasserman_optical_2025}. Examples of spectra for various system parameters are shown in Figure \ref{fig:e_EM_ep}.

\subsection{Proton Acceleration}\label{subsec:proton_acceleration}

As only the CLS accelerates particles, producing a CR energy density $\epsilon_{\rm CR}\rho v_{\rm CLS}^2/2$, the shock power that is deposited in CRs is
\begin{equation}
    L_{\rm CR}\left(r_{\rm sh}\right)=\epsilon_{\rm CR}\tilde{v}_{\rm CLS}\left(r_{\rm sh}\right)^2L_{\rm sh},
\end{equation}
where $\tilde{v}_{\rm CLS}$ is given by Equation (\ref{eq:v_CLS}). It is evident that if the CSM is ``too compact", as $\tau_{\rm e}$ approaches $3.3\,c/v$ (or $R_{\rm CLS}$ approaches $R_{\rm e}$), the energy deposited in CRs vanishes. We stress that it is important to properly take into account the fact that $v_{\rm CLS}$ is smaller than $v$ to compute the onset of CR acceleration and early light curves of neutrinos \citep[see also][]{tsuna_radiative_2023}.

\begin{figure}
    \centering
    \includegraphics[width=8.5cm]{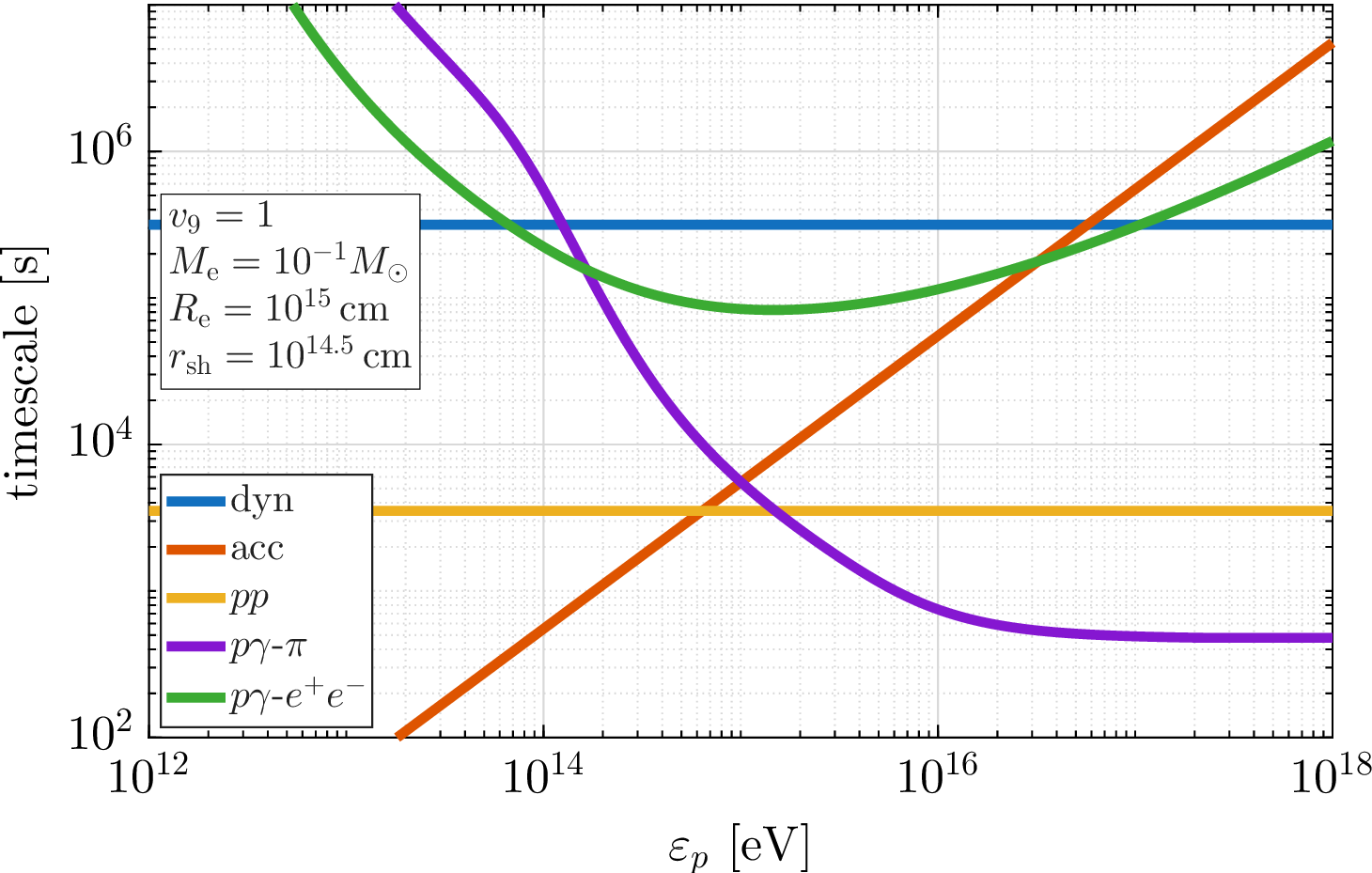}
    \caption{The dynamical timescale, and the proton acceleration and cooling timescales described in the text as a function of the proton energy. We use $\kappa_{\rm T}=0.34\,{\rm cm^2\,g^{-1}}$, $\epsilon_{\rm CR}=10^{-1}$ and $\epsilon_B=10^{-2}$.}
    \label{fig:p_timescales}
\end{figure}

Figure \ref{fig:p_timescales} shows the different timescales of energy gain and loss of the accelerated protons, as a function of the proton energy, which are described below.
The proton acceleration time in the Bohm limit is 
\begin{equation}
\begin{split}
    &t_{\rm acc}\left(r_{\rm sh},\varepsilon_p\right)=\frac{20}{3}\frac{c\varepsilon_p}{ev_{\rm CLS}^2B}=\frac{r_{\rm sh}}{v}\frac{\varepsilon_p}{10^{17}\,{\rm eV}}\tilde{v}_{\rm CLS}^{-3}\\
    &\times v_9^{-2}\epsilon_{B,-2}^{\frac{1}{2}}\left(\frac{M_{\rm e}}{10^{-1}\,M_\odot}\right)^{-\frac{1}{2}}\left(\frac{R_{\rm e}}{10^{15}\,\rm cm}\right)^{\frac{1}{2}},
\end{split}
\end{equation}
where $\varepsilon_p$ is the proton energy, $e$ is the electron charge, and the magnetic field is
\begin{equation}
    \label{eq:B}
    \begin{split}
    &B\left(r_{\rm sh}\right)=\sqrt{8\pi\epsilon_B\frac{2}{\gamma^2-1}\rho v_{\rm CLS}^2}\\
    &=53\,{\rm G}\,\frac{R_{\rm e}}{r_{\rm sh}}\tilde{v}_{\rm CLS}\epsilon_{B,-2}^{1/2}v_9\left(\frac{\rho_{\rm e}}{10^{-14}\,{\rm g\,cm^{-3}}}\right)^{1/2},
    \end{split}
\end{equation} 
with $\epsilon_{B,-2}=\epsilon_{B}/10^{-2}$.

We focus on times when the proton cooling timescale is shorter than the dynamical time, so that all the CR energy is dissipated efficiently, and neglect the additional dissipated energy at later times (see discussion below Equation (\ref{eq:R_pp})). As the dynamical time, which is the time available for acceleration, is comparable to or shorter than the proton escape and adiabatic suppression timescales, these do not play a role during the efficient cooling regime. In addition, in this regime, the protons can be assumed to lose their energy in a plasma density and radiation energy density and spectrum corresponding to the conditions at the immediate downstream \citep[the radiation density and spectrum are, in any case, quite uniform in the downstream vicinity, see][]{wasserman_optical_2025}.

As shown in Figure \ref{fig:p_timescales}, the dominant proton cooling process is $pp$ pion-production, and for higher proton energies it may be $p\gamma$ pion-production. We approximate the cross-sections as energy-independent above threshold, and the $p\gamma$ as having a collision angle $\theta=\pi$
\begin{equation}\label{eq:t_pp}
\begin{split}
    &t_{pp}\left(r_{\rm sh}\right)=\left(\frac{\gamma+1}{\gamma-1}\frac{\rho}{m_p}\hat{\sigma}_{pp}c\right)^{-1}\\
&=3.5\times 10^{-2}\frac{r_{\rm sh}}{v}\frac{r_{\rm sh}}{R_{\rm e}}v_9\left(\frac{M_{\rm e}}{10^{-1}\,M_\odot}\right)^{-1}\left(\frac{R_{\rm e}}{10^{15}\,\rm cm}\right)^2,
    \end{split}
\end{equation}
\begin{equation}\label{eq:t_pg}
\begin{split}
    &t_{p\gamma}^\pi\left(r_{\rm sh},\varepsilon_p\right)=\left(n_{\gamma}^\pi\hat{\sigma}_{p\gamma}c\right)^{-1},\\
&n_{\gamma}^\pi\left(r_{\rm sh},\varepsilon_p\right)\equiv\int_{m_{\pi}m_pc^4/\left(2\,\varepsilon_p\right)}^{\infty}d\varepsilon_\gamma\frac{e_{{\rm EM},\varepsilon_\gamma}}{\varepsilon_\gamma},
    \end{split}
\end{equation}
assuming $n_p=\rho/m_p$ and using ``effective'' cross-section values (including the inelasticity) of $\hat{\sigma}_{pp}=25\,{\rm mb}$, $\hat{\sigma}_{p\gamma}^\pi=0.1\,{\rm mb}$ and $m_\pi c^2=150\,{\rm MeV}$, such that $t_{pp}$ is independent of the proton energy. The time-dependent radiation energy spectrum at the shock is obtained from the numerical solution, as explained in the previous subsection, using Equation (\ref{eq:e_EM}) to normalize the radiation energy density. The $p\gamma$ pair-production energy loss is also shown in Figure \ref{fig:p_timescales} \citep[using][]{chodorowski_reaction_1992}, and it is always subdominant in our parameter range (as $\hat{\sigma}_{p\gamma}^{e^+e^-}\approx10^{-3.5}\,{\rm mb}$).

The radius at which the $pp$ time is equal to the dynamical time is \citep{katz_x-rays_2011}\footnote{This corresponds to the radius at which $f_{pp}=t_{\rm dyn}/t_{pp}=1$ \citep{murase_new_2011,murase_new_2018}. We assume CRs interact under fixed immediate downstream conditions, although CR transport through the downstream, the time-dependent adiabatic index, and escape/adiabatic losses may introduce some correction.}
\begin{equation}
\label{eq:R_pp}
    R_{pp}=4\frac{\hat{\sigma}_{pp}}{{\kappa}_{\rm T}m_p}\frac{c}{v} \tau_{\rm e}R_{\rm e}=160\,R_{\rm bo}^\infty\kappa_{0.34}^{-1}v_9^{-2}.
\end{equation}  
Thus, a $pp$-inefficiency radial region exists for sufficiently extended CSM of $R_{\rm e}>5.3\times 10^{15}\,{\rm cm}\,v_9^{-1}\left(M_{\rm e}/10^{-1}\,M_\odot\right)^{1/2}$, or $\tau_{\rm e}/\left(c/v\right)<6.3\times10^{-3}\,\kappa_{0.34}v_9^{2}$. $R_{pp}$ is beyond $R_{\rm ad}$, Equation (\ref{eq:R_ad}), for $0.1<v_9<1.6$, see example in Figure \ref{fig:profiles} and Figure \ref{fig:tau_gg}.

We neglect the additional energy dissipated above $R_{pp}$ in our treatment. In principle, $t_{\rm dyn}/t_{pp}\propto r_{\rm sh}^{-1}$ implies a logarithmic increase in the total energy, but as in reality $v\left(r_{\rm sh}\right)$ is slowly declining (Section \ref{sec:intro}), the dissipated energy is dominated by the $pp$-efficient radial region. The $p\gamma$ contribution above $R_{pp}$ is also not significant as it is relevant only for the most energetic protons. We thus track the emission up to 
\begin{equation}
    \label{eq:R_lim}
    R_{\rm lim}\equiv\min\left(R_{\rm e},R_{pp}\right).
\end{equation}

The maximal proton energy at each shock radius is thus roughly given by equating the acceleration time with the cooling time
\begin{equation}
\label{eq:ep_max}
\begin{split}
    t_{\rm acc}=&\min\left(t_{pp},t_{p\gamma}\right)\overset{t_{pp}<t_{p\gamma}}{\implies}\\
    \varepsilon_{p}^{\rm max}\left(r_{\rm sh}\right)
    =&4.4\,{\rm PeV}\frac{r_{\rm sh}}{R_{\rm e}}\tilde{v}_{\rm CLS}^3\\
    &\times v_9^3\epsilon_{B,-2}^{\frac{1}{2}}\left(\frac{\rho_{\rm e}}{10^{-14}\,{\rm g\,cm^{-3}}}\right)^{-\frac{1}{2}}.
\end{split}
\end{equation}
This gives the $pp$-limited maximal energy \citep{murase_new_2011}. When $p\gamma$ losses dominate, $\varepsilon_p^{\max}$ is obtained numerically from $t_{\rm acc}=t_{p\gamma}$ using the radiation spectrum as described above.

\subsection{Neutrino Production}\label{subsec:nu_prod}

We assume a flat power-law energy distribution for the accelerated protons from $1\,$GeV\footnote{The maximal proton energy crosses this threshold close to $R_{\rm CLS}$.} up to $\varepsilon_p^{\max}$, and as the typical neutrino energy is $\sim0.03-0.05$ of the parent proton energy, we use a fixed fraction $\varepsilon_\nu=0.05\,\varepsilon_p$.

\begin{figure}
    \centering
    \includegraphics[width=8.5cm]{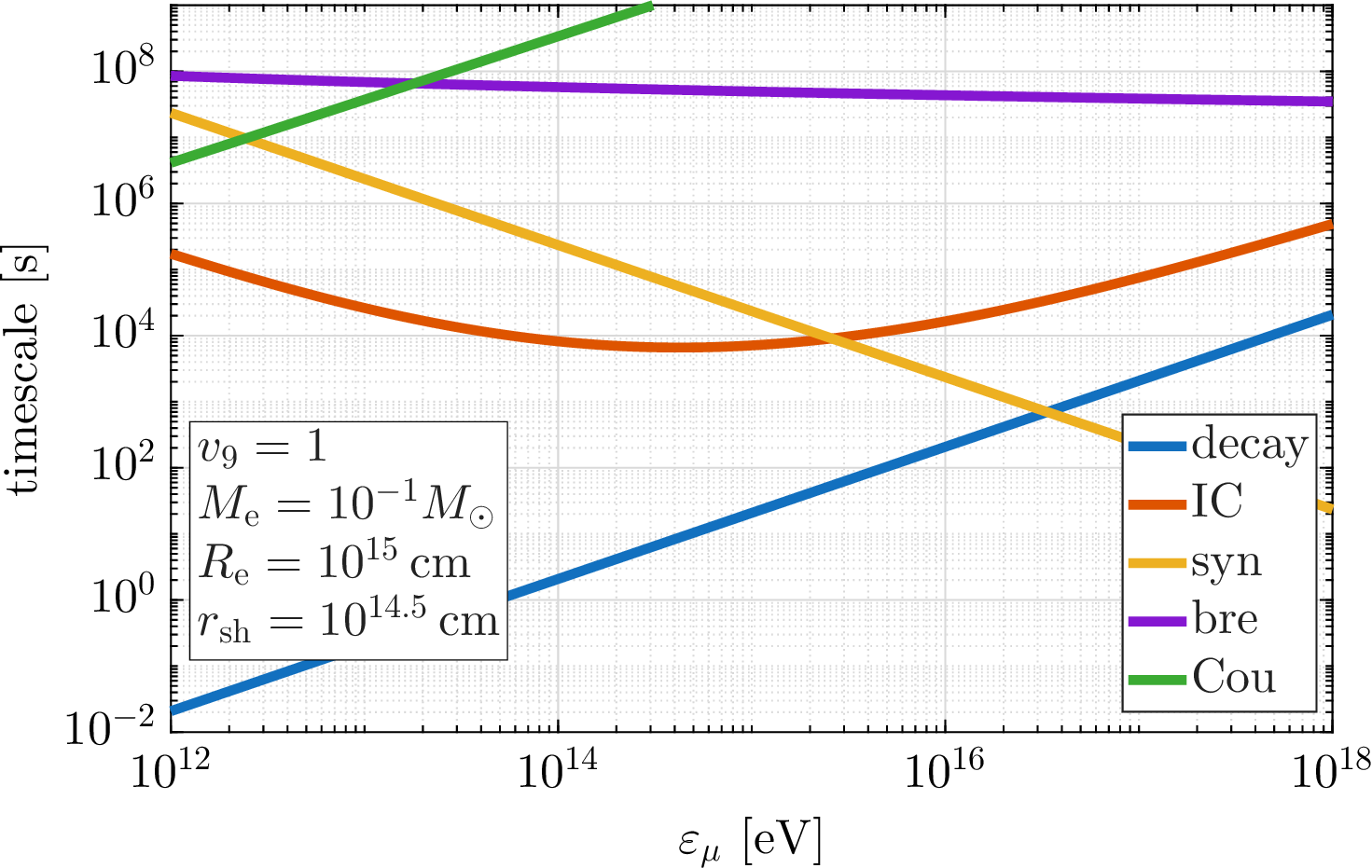}
    \caption{The muon decay timescale, and the muon cooling timescales by inverse-Compton (using the ultra-relativistic (UR), $\theta=\pi$, $\delta$-approximation with Klein-Nishina suppression), by UR synchrotron, by bremsstrahlung (using the UR weak-shielding Bethe-Heitler loss rate, including proton and electron targets), and by Coulomb collisions (for UR massive particle in ionized plasma), as a function of the muon energy \citep{blumenthal_bremsstrahlung_1970,jackson_classical_1998}. We use $\kappa_{\rm T}=0.34\,{\rm cm^2\,g^{-1}}$, $\epsilon_{\rm CR}=10^{-1}$ and $\epsilon_B=10^{-2}$.}
    \label{fig:mu_time}
\end{figure}

In the multi-pion-production regime, the lost proton energy is roughly distributed between different pion charges as $\pi^+:\pi^-:\pi^0\sim1:1:1$. The charged pions decay to $\pi^\pm\rightarrow\mu^\pm+\nu_\mu\left(\bar{\nu}_\mu\right)$ (while the neutral pion to $\gamma+\gamma$), and the muons then further decay to $\mu^\pm\rightarrow e^\pm+\nu_e\left(\bar\nu_e\right)+\bar\nu_\mu\left(\nu_\mu\right)$. Figure \ref{fig:mu_time} shows different energy loss timescales for the muons as a function of energy, showing that they do not suffer significant losses before decaying (for $\varepsilon_\mu\lesssim10^{16}\,{\rm eV}$, in all of our parameter range). We thus treat the decay of muons and the pions (which decay even faster) as instantaneous.

As the three neutrinos carry about $3/4$ of the parent charged-pion energy, the total fraction\footnote{We note that some of the CR protons are converted by pion-production to neutrons, which escape the system quickly on $r_{\rm sh}/c$ timescale since they are not magnetically confined, and since their decay time is longer than $r_{\rm sh}/c$ for energies $\varepsilon_n>1\,{\rm TeV}\left(r_{\rm sh}/10^{16.5}\,{\rm cm}\right)$. We thus expect that at radii $r_{\rm sh}>\left(v/c\right)R_{pp}=5.3\,R_{\rm bo}^\infty\kappa_{0.34}^{-1}v_9^{-1}$, escaping neutrons will carry away some fraction of the energy instead of converting it to pions in pion-production. Assuming that the inelasticity is $\kappa=0.5$ and that on average $1/3$ of the protons are converted to neutrons in each interaction, the neutrons are expected to escape with about $0.2$ of the initial energy at radii above the mentioned radius. We neglect this correction in our treatment.} of energy converted to all-flavor neutrinos (including anti-neutrinos) is $2/3\times3/4$, hence in the efficient CR cooling region
\begin{equation}
\label{eq:L_nu}
    L_\nu\left(r_{\rm sh}\right) 
    \approx \frac{1}{2} \,L_{\rm CR}\left(r_{\rm sh}\right) ,
\end{equation}
where half of this power is carried by neutrinos and half by anti-neutrinos. Neglecting the varying light travel time due to shock radius propagation (as $v\ll c$), the observed neutrino luminosity is $L_\nu^{\rm obs}\left(t\right) \approx L_\nu\left(r_{\rm sh}=vt\right)$.

As the charged pions are produced in equal amounts (and as the pions and muons do not suffer significant losses before decaying), the neutrino flavor ratio produced at the source is $\nu_e:\nu_\mu:\nu_\tau=1:2:0$ (for both neutrinos and anti-neutrinos), such that after propagation, the ratio at Earth is close to $1:1:1$. In $p\gamma$ pion-production near the $\Delta$ resonance, a single pion is produced, such that the $\pi^-$ channel is not allowed by charge conservation, and the pion ratio becomes approximately $\pi^+:\pi^-:\pi^0\sim1:0:1$ due to the contribution from direct pion production. This pion ratio implies that the corresponding neutrino power (see Equation \ref{eq:L_nu}), is reduced by a factor $3/4$ relative to the multi-pion regime, and is split between neutrinos and anti-neutrinos as $2:1$ instead of by equal amount, and that the neutrino flavor produced at the source is $1:1:0$ and $0:1:0$, such that the ratio at Earth is approximately $1:0.8:0.8$ and $1:1.9:1.8$, for neutrinos and anti-neutrinos respectively \citep[using PMNS matrix from][]{esteban_nufit-60_2024}. We thus expect a transition of the flavor ratio from lower energy neutrinos produced in $pp$ pion-production to higher energy neutrinos produced in $p\gamma$ pion-production through the $\Delta$ resonance. Measurements of the transition energy and energy width will provide unique probes of the sources' physics and direct constraints on the plasma density and radiation spectrum \citep[see][]{kashti_astrophysical_2005}. In what follows, however, we estimate only the all-flavor neutrino (and anti-neutrino) signal.

For the flat CR spectrum, the all-flavor neutrino (including anti-neutrino) production rate is
\begin{equation}
    \varepsilon_\nu^2\frac{d\dot{n}_\nu}{d\varepsilon_\nu}\left(r_{\rm sh}\right)=\frac{L_\nu}{\ln\frac{\varepsilon_p^{\max}}{{\rm GeV}}}
\end{equation}
for $0.05\,{\rm GeV}<\varepsilon_\nu<0.05\,\varepsilon_p^{\max}$.
The neutrino power above $1\,$TeV, for example, is
\begin{equation}
\label{eq:L_nu_TeV}
    L_\nu^{>1\,{\rm TeV}}\left(r_{\rm sh}\right)=\frac{\ln\frac{\varepsilon_\nu^{\max}}{{\rm TeV}}}{\ln\frac{\varepsilon_p^{\max}}{{\rm GeV}}}L_\nu,
\end{equation}
where $\varepsilon_\nu^{\max}\equiv0.05\,\varepsilon_p^{\max}$ (assumed larger than $1$ TeV).

\section{Neutrino Light Curves \& Spectra}\label{sec:nu_emission}

\begin{figure}
    \centering
    \includegraphics[width=8.5cm]{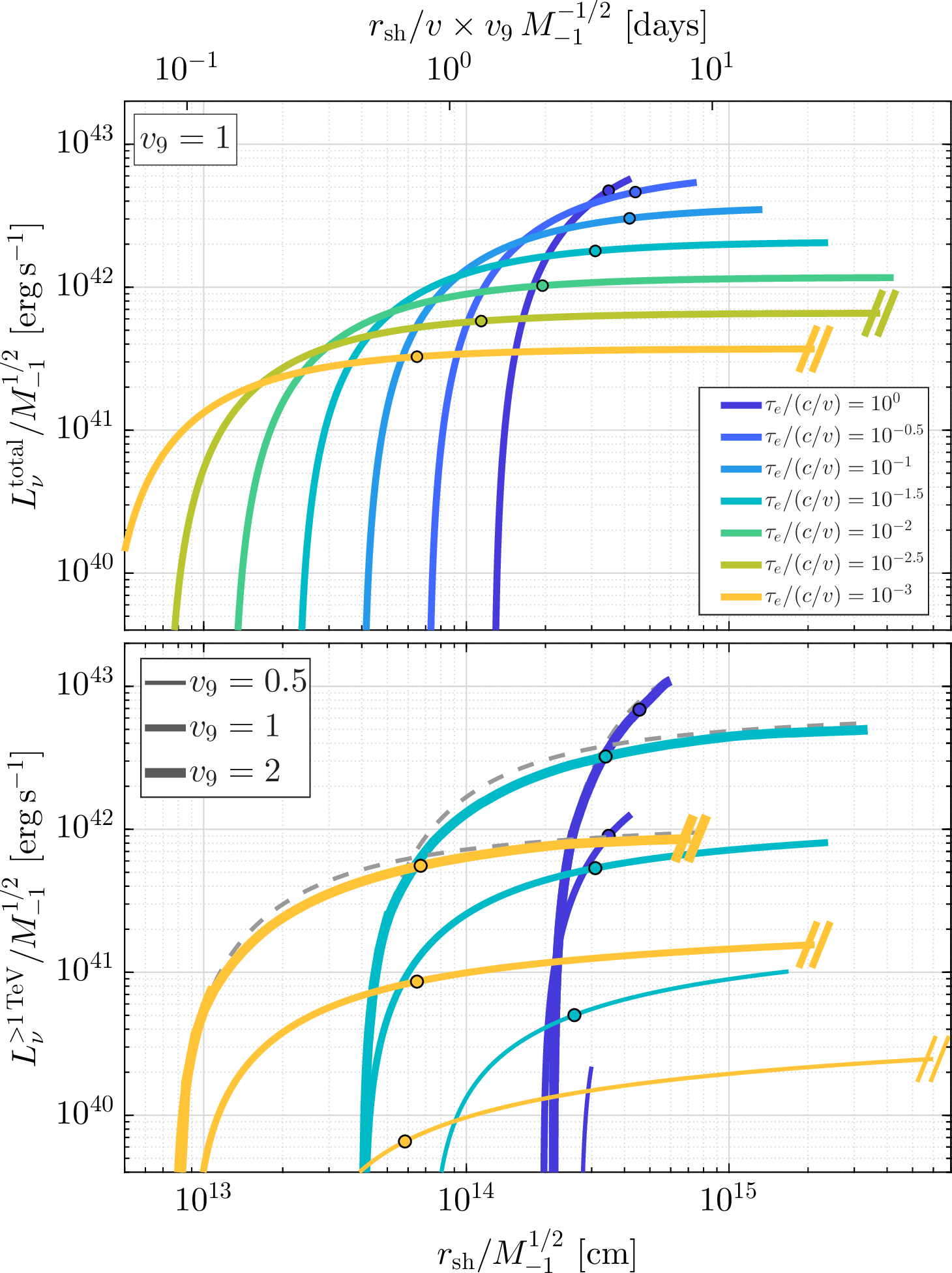}
    \caption{The neutrino luminosity as a function of shock radius (or time, shown in the upper $x$ axis), Equation (\ref{eq:L_nu_TeV}), for different characteristic CSM optical depths $\tau_{\rm e}/\left(c/v\right)$, defined in Equation (\ref{eq:tau_e}). The upper panel shows the total luminosity for $v_9=1$, and the lower panel shows the luminosity above $1\,$TeV for different $v_9$ values. The gray dashed line shows the $pp$-only light curve neglecting $p\gamma$ interactions, the contribution of which is apparent only for the $v_9=2$ case. The axes are normalized by the $M_{\rm e}$ dependence (for a fixed $\tau_{\rm e}/\left(c/v\right)$), such that the upper panel is exact for all $M_{\rm e}$, while the lower panel shows results for $M_{\rm e}=10^{-1}\,M_\odot$, and the variations within an order of magnitude in $M_{\rm e}$ change the curves by an amount comparable to the line width (through the weak dependence of the maximal energy, see Section \ref{subsec:proton_acceleration}). The light curves are shown up to $R_{\rm lim}$, Equation (\ref{eq:R_lim}), which is the minimum between the CSM radius $R_{\rm e}$ and the $pp$-inefficiency radius $R_{pp}$. The dots mark the corresponding peak time of the breakout EM emission, as given by Equation (\ref{eq:t_EM}). We use $\kappa_{\rm T}=0.34\,{\rm cm^2\,g^{-1}}$, $\epsilon_{\rm CR}=10^{-1}$ and $\epsilon_B=10^{-2}$.}
    \label{fig:L_nu_tot} 
\end{figure}

\begin{figure}
    \centering
    \includegraphics[width=8.5cm]{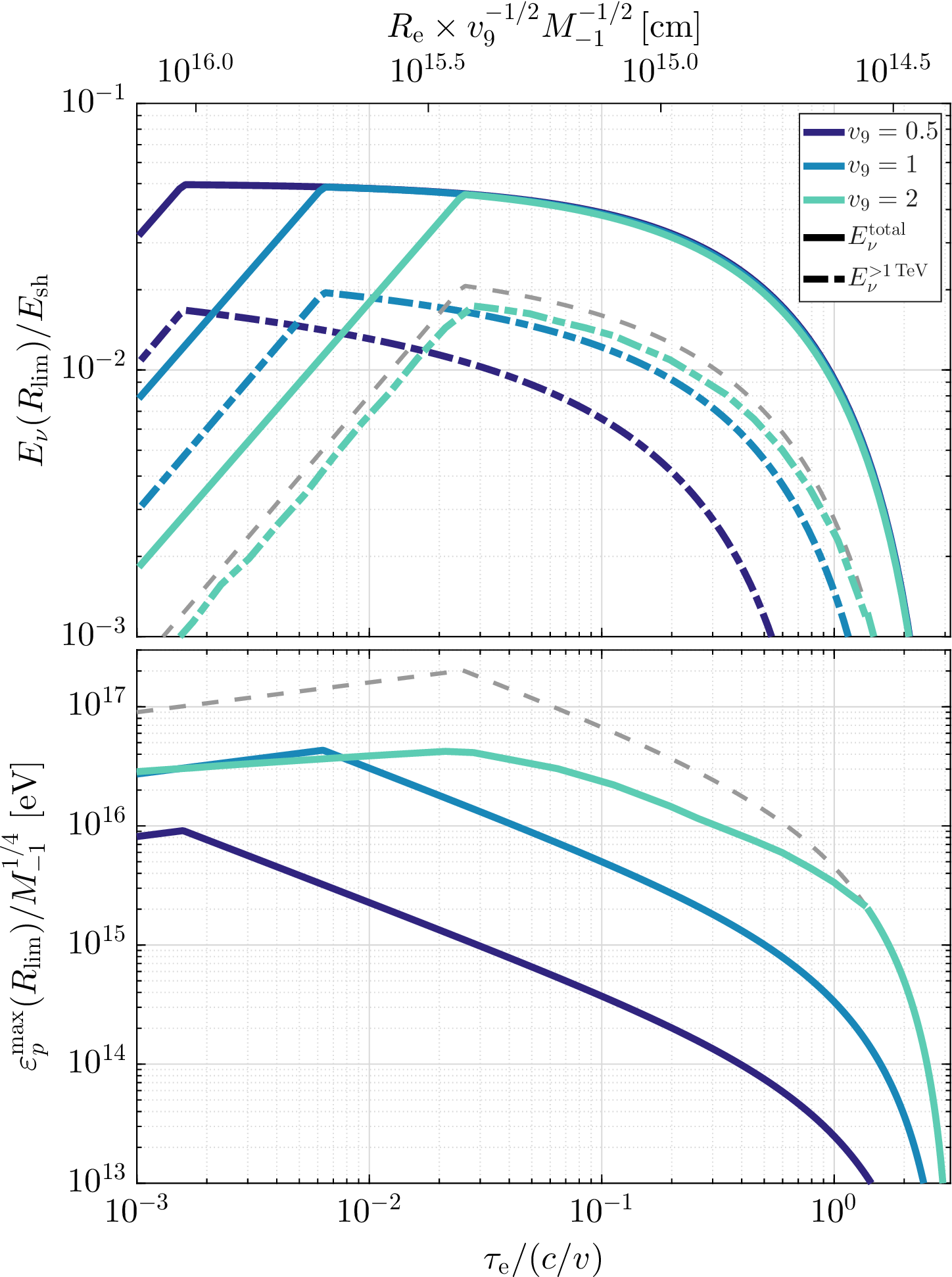}
    \caption{The upper panel shows the integrated neutrino energy emitted up to $R_{\rm lim}$ as a function of characteristic CSM optical depth, for different $v_9$, from Equations (\ref{eq:E_nu_TeV})-(\ref{eq:E_nu}). Both the total neutrino energy and the $>1\,$TeV energy are shown, normalized to the total shock energy $E_{\rm sh}$, Equation (\ref{eq:S(r)}). The lower panel shows the maximal accelerated proton energy at $R_{\rm lim}$ as a function of characteristic CSM optical depth for different $v_9$, Equation (\ref{eq:ep_max}). The gray dashed line shows the $pp$-only result neglecting $p\gamma$ interactions, the contribution of which is apparent only for $v_9=2$. The upper $x$ axis shows the corresponding $R_{\rm e}$ values. The axes are normalized such that the total integrated energy and the maximal proton energy neglecting $p\gamma$ are exact for all $M_{\rm e}$, while the $>1\,$TeV energy and the $p\gamma$ corrections are shown for $M_{\rm e}=10^{-1}\,M_\odot$, and variations within an order of magnitude in $M_{\rm e}$ change the curves by an amount comparable to the line width. We use $\kappa_{\rm T}=0.34\,{\rm cm^2\,g^{-1}}$, $\epsilon_{\rm CR}=10^{-1}$ and $\epsilon_B=10^{-2}$.}
    \label{fig:E_nu_tot}
\end{figure}

\begin{figure}
    \centering
    \includegraphics[width=8.5cm]{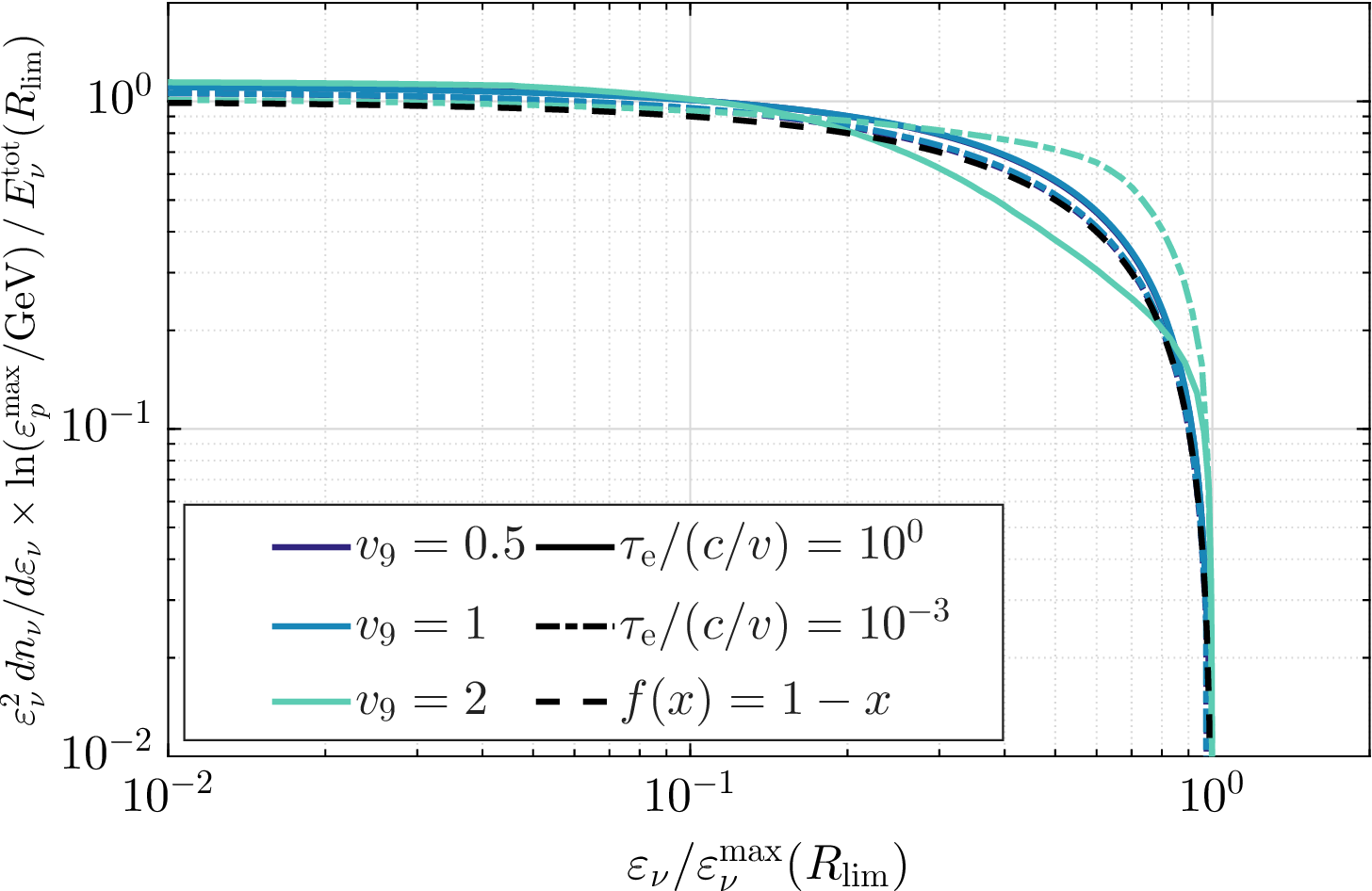}
    \caption{The neutrino energy distribution accumulated up to $R_{\rm lim}$, from Equation (\ref{eq:nu_spect}), normalized according to the pre-factor of Equation (\ref{eq:spectrum_approx2}). Results are presented for different $v_9$ and $\tau_{\rm e}/\left(c/v\right)$ values. The $f\left(x\right)=1-x$ approximation (with $x\equiv\varepsilon_\nu/\varepsilon_\nu^{\rm max}$), Equation (\ref{eq:spectrum_approx2}), provides a good description of the spectrum for all cases with physical parameters in the range considered (including different CSM masses, although only $M_{\rm e}=10^{-1}M_\odot$ is presented). We use $\kappa_{\rm T}=0.34\,{\rm cm^2\,g^{-1}}$, $\epsilon_{\rm CR}=10^{-1}$ and $\epsilon_B=10^{-2}$.}
    \label{fig:nu_spect}
\end{figure}

Figure \ref{fig:L_nu_tot} shows the neutrino luminosity as a function of shock radius or time, and Figure \ref{fig:E_nu_tot} shows the total emitted neutrino energy, and the maximal accelerated proton energy, as a function of the physical parameters and the characteristic CSM optical depth, $\tau_{\rm e}/\left(c/v\right)$, defined in Equation (\ref{eq:tau_e}). Figure \ref{fig:nu_spect} shows the total emitted neutrino energy distribution.

The emitted neutrino energy up to radius $r_{\rm sh}$ can be obtained by integrating the luminosity, Equation (\ref{eq:L_nu_TeV})
\begin{equation}
\label{eq:E_nu_TeV}
    E_\nu^{>1\,{\rm TeV}}\left(r_{\rm sh}\right)=\int_{R_{\rm CLS}}^{r_{\rm sh}}\frac{dr}{v}L_\nu^{>1\,{\rm TeV}}\left(r\right).
\end{equation}
The total energy integral can be solved analytically
\begin{equation}
\label{eq:E_nu}
\begin{split}
    E_\nu^{\rm tot}\left(r_{\rm sh}\right)&=\frac{1}{2}\,\epsilon_{\rm CR}\frac{L_{\rm sh}}{v}\int_{R_{\rm CLS}}^{r_{\rm sh}}dr\left(1-\frac{R_{\rm CLS}}{r}\right)^2\\
    &=\frac{1}{2}\,\epsilon_{\rm CR}E_{\rm sh}\left(R_{\rm e}\right)f\left(r_{\rm sh}\right),
\end{split}
\end{equation}
where
\begin{equation}
\label{eq:S(r)}
\begin{split}
    &E_{\rm sh}\left(r_{\rm sh}\right)\equiv L_{\rm sh}\times \frac{r_{\rm sh}}{v}=10^{50}\,{\rm erg}\frac{r_{\rm sh}}{R_{\rm e}}M_{-1}v_9^2,\\
    &f\left(r_{\rm sh}\right)\equiv\frac{r_{\rm sh}}{R_{\rm e}}\left[1+2\frac{R_{\rm CLS}}{r_{\rm sh}}\ln \frac{R_{\rm CLS}}{r_{\rm sh}}-\left(\frac{R_{\rm CLS}}{r_{\rm sh}}\right)^2\right],
\end{split}
\end{equation}
with $M_{\rm e}=10^{-1}\,M_\odot\,M_{-1}$. Evaluating the neutrino energy emitted up to $R_{\rm lim}$, defined in Equation (\ref{eq:R_lim}), which is the minimum of the CSM radius $R_{\rm e}$ and the $pp$-inefficiency radius $R_{pp}$, we obtain both the suppression at $\tau_{\rm e}/\left(c/v\right)\rightarrow3.3$, where the CLS onset is close to the CSM edge $R_{\rm e}$, and the suppression at low $\tau_{\rm e}/\left(c/v\right)\ll1$, where an increasing fraction of the CSM mass is at a density too low for an efficient $pp$ pion-production, see the suppression function $f\left(R_{\rm lim};\tau_{\rm e},v\right)$ from Equations (\ref{eq:E_nu})-(\ref{eq:S(r)}) and Figure \ref{fig:E_nu_tot}. The neutrino energy emitted above some energy, $1$ TeV, for example, cannot be written in a closed form, but a convenient approximation with $\lesssim 50\%$ error is obtained by setting the maximal proton energy to be the one achieved at $R_{\rm lim}$, yielding
\begin{equation}
    E_\nu^{>1\,{\rm TeV}}\left(R_{\rm lim}\right)\approx \frac{\ln\frac{\varepsilon_\nu^{\max}\left(R_{\rm lim}\right)}{{\rm TeV}}}{\ln\frac{\varepsilon_p^{\max}\left(R_{\rm lim}\right)}{{\rm GeV}}}\times E_\nu^{\rm tot}\left(R_{\rm lim}\right).
\end{equation}

The  maximal proton energy obtained is always limited to $\varepsilon_p^{\max}\lesssim20\,{\rm PeV}\,\epsilon_{B,-2}^{1/2}$ (corresponding to $\varepsilon_\nu^{\max}\lesssim1\,{\rm PeV}\,\epsilon_{B,-2}^{1/2}$). The $\epsilon_B$ scaling is also correct for the $p\gamma$-limited maximal energy (which is obtained at $v_9>1$), as at $\sim10$ PeV energies the proton can already undergo $p\gamma$-interactions with most of the background photons. This limit should also remain an upper bound for $M_{\rm e}>1\,M_\odot$, where the shock decelerates significantly.

Similarly, the accumulated energy distribution is
\begin{equation}
    \label{eq:nu_spect}
    \varepsilon_\nu^2\frac{dn_\nu}{d\varepsilon_\nu}\left(\varepsilon_\nu,r_{\rm sh}\right)=\int_{R_*\left(\varepsilon_\nu\right)}^{r_{\rm sh}}\frac{dr}{v}\frac{L_\nu\left(r\right)}{\ln\frac{\varepsilon_p^{\max}\left(r\right)}{{\rm GeV}}},
\end{equation}
where $R_*\left(\varepsilon_\nu\right)$ is defined as the solution of $\varepsilon_\nu^{\max}\left(r\right)=\varepsilon_\nu$. Denoting $x=\varepsilon_\nu/\varepsilon_\nu^{\max}\left(r_{\rm sh}\right)$, a very good approximation for this expression is
\begin{equation}
\label{eq:spectrum_approx1}
    \varepsilon_\nu^2\frac{dn_\nu}{d\varepsilon_\nu}\left(x,r_{\rm sh}\right)\approx E_\nu^{\rm tot}\left(r_{\rm sh}\right)\int_{x}^1\frac{d\xi}{\ln\left(\frac{\varepsilon_p^{\max}\left(r_{\rm sh}\right)}{\rm GeV}\xi\right)},
\end{equation}
where for $\Lambda\equiv\varepsilon_p^{\max}\left(r_{\rm sh}\right)/{\rm GeV}\gg1$ the logarithmic integral function simplifies to ${\rm li}\,{\Lambda}\rightarrow\Lambda/\ln \Lambda$, such that the particular integral simplifies to
\begin{equation}
\label{eq:1-x}
    \frac{{\rm li}\,\Lambda-{\rm li}\,\left(\Lambda x\right)}{\Lambda}\approx \frac{1}{\ln{\Lambda}}\left[1-\frac{\ln{\Lambda}}{\ln{\left(\Lambda x\right)}}x\right]\approx\frac{1}{\ln{\Lambda}}\left(1-x\right),
\end{equation}
and
\begin{equation}
    \label{eq:spectrum_approx2}
    \varepsilon_\nu^2\frac{dn_\nu}{d\varepsilon_\nu}\left(x,r_{\rm sh}\right)\approx \frac{E_\nu^{\rm tot}\left(r_{\rm sh}\right)}{\ln{\Lambda}}\left(1-x\right),
\end{equation}
providing a simple and accurate description of the accumulated neutrino energy distribution for different physical parameters, see Figure \ref{fig:nu_spect}.

For a fixed total CR energy, a proton spectrum $dn_p/d\varepsilon_p \propto \varepsilon_p^{-s}$ with $s>2$ yields a high-energy neutrino flux that is lower than that obtained for an $s=2$ spectrum because a larger fraction of the energy is carried by low-energy protons. Away from the high-energy cutoff, the neutrino energy distribution scales as $\varepsilon_\nu^2dn_\nu/d\varepsilon_\nu\propto\varepsilon_\nu^{2-s}$. For a proton spectrum with $s>2$, Equation (\ref{eq:spectrum_approx2}) should be multiplied by
\begin{equation}
\label{eq:spectrum_approx_s}
    {\cal S}_s\left(\varepsilon_\nu\right)
    \equiv
    \ln\Lambda\,
    \frac{s-2}
    {
        1-
        \Lambda^{2-s}
    }
    \left(
        \frac{\varepsilon_\nu}
        {0.05\,{\rm GeV}}
    \right)^{2-s}.
\end{equation}

This expression connects continuously to the flat proton spectrum result, with $\lim_{s\rightarrow2^+}{\cal S}_s(\varepsilon_\nu)=1$. For $s=2.2$ and $\varepsilon_p^{\max}=3$ PeV, we obtain ${\cal S}_s\left(10\,{\rm TeV}\right)\approx0.3$ \citep[see also][]{murase_new_2018}.

\section{Contribution to the Neutrino Background }\label{sec:nu_background}

\begin{figure}
    \centering
    \includegraphics[width=8.5cm]{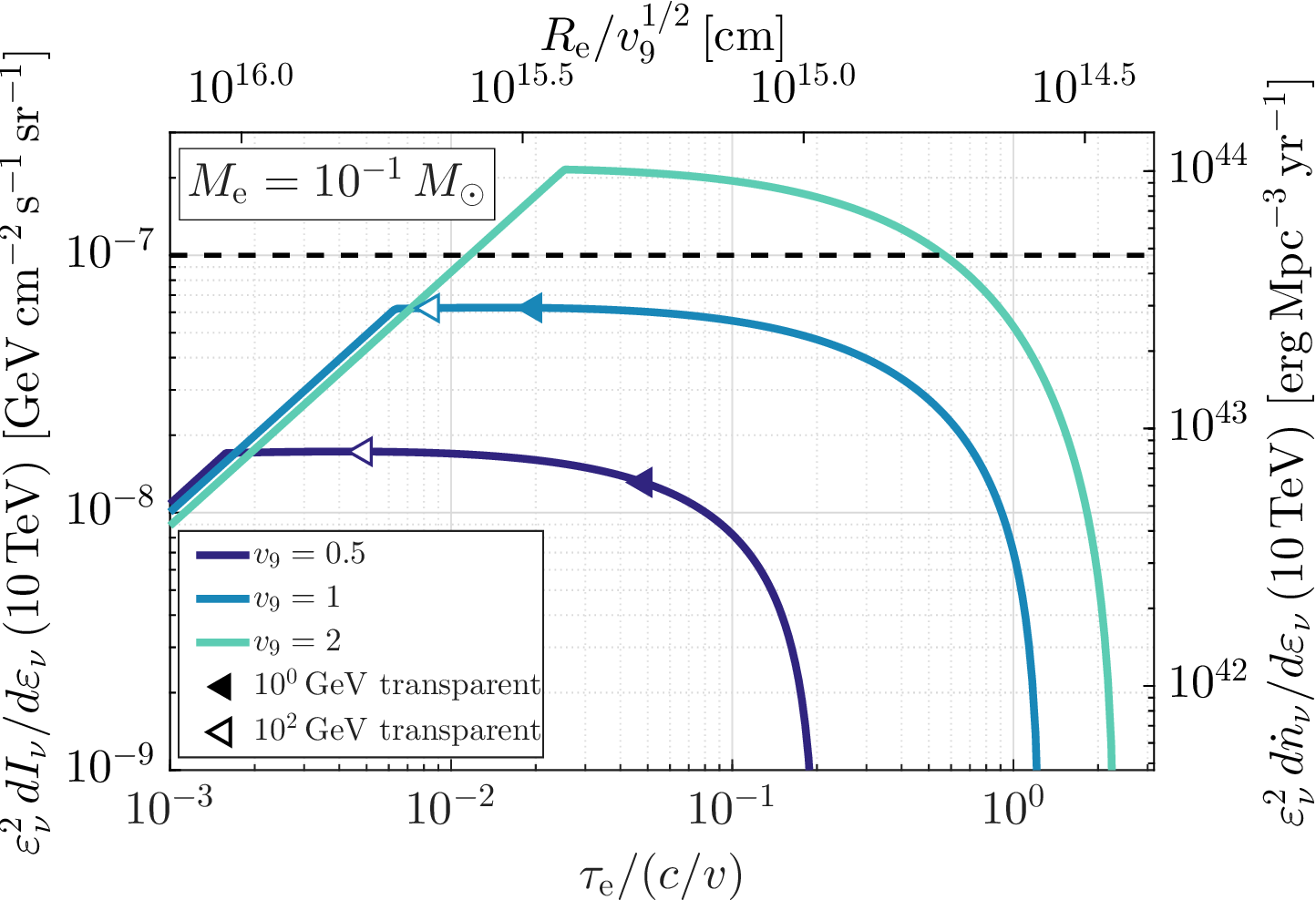}
    \caption{The $10$ TeV (all-flavor) neutrino flux at Earth, for a local rate of $R_{\rm SN}=10^{-4}\,{\rm Mpc^{-3}\,yr^{-1}}$ identical sources, as a function of their characteristic CSM optical depth value, for different $v_9$ and for $M_{\rm e}=10^{-1}\,M_\odot$, from Equation (\ref{eq:nu_flux_xiz}). The upper $x$ axis shows the corresponding $R_{\rm e}$ values, and the right $y$ axis shows the corresponding local energy production rate. The flux is approximately linear in $M_{\rm e}$; see Equation (\ref{eq:flux_approx}). The horizontal dashed line presents the observed value. The filled (open) triangle symbols present the characteristic CSM optical depth value, below which the system is transparent to $1(100)$ GeV $\gamma\gamma$ pair-production, such that a comparable flux of gamma rays is expected to be emitted, see Section \ref{sec:g_supp}. We use $\kappa_{\rm T}=0.34\,{\rm cm^2\,g^{-1}}$, $\epsilon_{\rm CR}=10^{-1}$ and $\epsilon_B=10^{-2}$.}
    \label{fig:flux_10TeV}
\end{figure}

Figure \ref{fig:flux_10TeV} shows the $10$ TeV neutrino flux obtained at Earth, for a local rate of $R_{\rm SN}/10^{-4}\,{\rm Mpc^{-3}\,yr^{-1}}\equiv R_{-4}$ identical sources, as a function of their characteristic CSM optical depth $\tau_{\rm e}/\left(c/v\right)$ value.

The neutrino flux at Earth is given by
\begin{equation}
\label{eq:nu_flux}
    \frac{dI_\nu}{d\varepsilon_\nu}\left(\varepsilon_\nu\right)=\frac{c}{4\pi}\int_0^{z_{\max}}dz\frac{Q_\nu\left[\left(1+z\right)\varepsilon_\nu,z\right]}{H\left(z\right)},
\end{equation}
where $Q_\nu$ is the neutrino number production rate, and $H$ is the Hubble parameter. For a fixed power law distribution, the production rate is a separable function of redshift and energy, such that for a flat distribution
\begin{equation}
\begin{split}
\label{eq:nu_flux_xiz}
    \frac{dI_\nu}{d\varepsilon_\nu}\left(\varepsilon_\nu\right)&=\frac{c t_H}{4\pi }Q_\nu\left(\varepsilon_\nu,0\right)\int_0^{z_{\max}}dz\frac{t_H^{-1}}{H\left(z\right)}\frac{g\left(z\right)}{\left(1+z\right)^2}\\
    &=\frac{c t_H}{4\pi}R_{\rm SN}\frac{dn_\nu\left(\varepsilon_\nu\right)}{d\varepsilon_\nu}\xi_z,
    \end{split}
\end{equation}
where $Q_\nu\left(\varepsilon_\nu,0\right)=R_{\rm SN}\times dn_\nu/d\varepsilon_\nu$, $t_H=\int_0^{\infty}dz\left|dt/dz\right|$ is the Hubble time, and $g\left(z\right)$ is the evolution of the event rate with redshift. The integral is denoted by $\xi_z$, a dimensionless parameter of order unity. We note that $\xi_z\approx 3$ is obtained for redshift evolution following that of the star formation rate or AGN luminosity density, $\propto\left(1 + z\right)^3$ up to $z = 2$ and constant at higher $z$. $\xi_z \approx 0.6$ is obtained for no evolution \citep[][]{waxman_high_1998}.

For neutrino energies well below the maximum energy, the last equality holds, justified because the spectrum is very close to a power law. We can also use the approximations of Equations (\ref{eq:spectrum_approx1})-(\ref{eq:spectrum_approx2}), to obtain
\begin{equation}
\label{eq:flux_approx}
    \begin{split}
    \varepsilon_\nu^2\frac{dI_\nu}{d\varepsilon_\nu}\sim &7\times 10^{-8}\frac{{\rm GeV}}{\rm cm^2\,s\,sr}R_{-4}\left(\frac{\epsilon_{\rm CR}}{10^{-1}}\right)M_{-1}v_9^2\\
    &\times \left(\frac{\ln \Lambda}{15}\right)^{-1}\left(\frac{\xi_z}{3}\right)f\left(R_{\rm lim}\right)\left(1-x\right),
    \end{split}
\end{equation}
where $\ln \Lambda$, $f\left(r_{\rm sh}\right)$, and $x$ were defined in the previous section. This closed-form approximation is very accurate, and it deviates from the exact integral result of Equation (\ref{eq:nu_flux_xiz}) and Figure \ref{fig:flux_10TeV} by the order of the curves' widths. For energies near the maximal neutrino energy, the spectrum decays and deviates from a power law, and the cosmological evolution will make this decay sharper in the local flux. We did not correct for this effect, which will sharpen somewhat only the rising part of the curves at large characteristic CSM optical depth in Figure \ref{fig:flux_10TeV}. The $p\gamma$ interactions, limiting the maximal neutrino energy, have a negligible impact on the accumulated neutrino flux at $10$ TeV, but may have a larger effect on the flux at higher neutrino energies (for $v_9>1$). For a nonidentical source distribution, the accumulated spectrum at Earth will not, in general, be a single power law, and the resulting flux should be convolved with the physical source parameters distribution, such that Equation (\ref{eq:flux_approx}) would be replaced by its population average (see discussion in Section \ref{sec:discussion}).

The fiducial rate $R_{\rm SN}=10^{-4}\,{\rm Mpc^{-3}\,yr^{-1}}$ is comparable to the total local core-collapse supernova rate rather than the rate of the hydrogen-rich source population. Hydrogen-poor SNe Ib/c and transitional SNe IIb are not, in general, described by the same fiducial CSM model and may be excluded from this rate, resulting in $R_{-4}\approx0.6-0.7$ correspondingly \citep{li_nearby_2011,smith_observed_2011}. If only a fraction of the hydrogen-rich progenitors possess the dense and compact CSM considered here, the effective source rate is further reduced.

For a proton spectrum with $s=2.2$ (Equation \ref{eq:spectrum_approx_s}) and with a rate $R_{-4}\approx 0.6$--$0.7$, the predicted $10$ TeV flux is $\sim0.2$ of the value derived above, for the same values of $M_{\rm e}$, $v$, and $\epsilon_{\rm CR}$. On the other hand, the rarer, $R_{-4}\approx 0.1$ IIn SN may contribute a similar fraction of the flux if they are indeed characterized by larger compact CSM masses. Given the uncertainties, the $s=2$, $R_{-4}=1$ result shown in Figure \ref{fig:flux_10TeV} and Equation (\ref{eq:flux_approx}) should be interpreted as indicating that compact CSM breakouts may contribute significantly to the background. Future EM observations of early SN emission will enable us to better constrain the prevalence and properties of dense CSM around SN progenitors, and the CSM breakout contribution to the neutrino background.

\section{Gamma-Ray Suppression}\label{sec:g_supp}

\begin{figure}
    \centering
    \includegraphics[width=8.5cm]{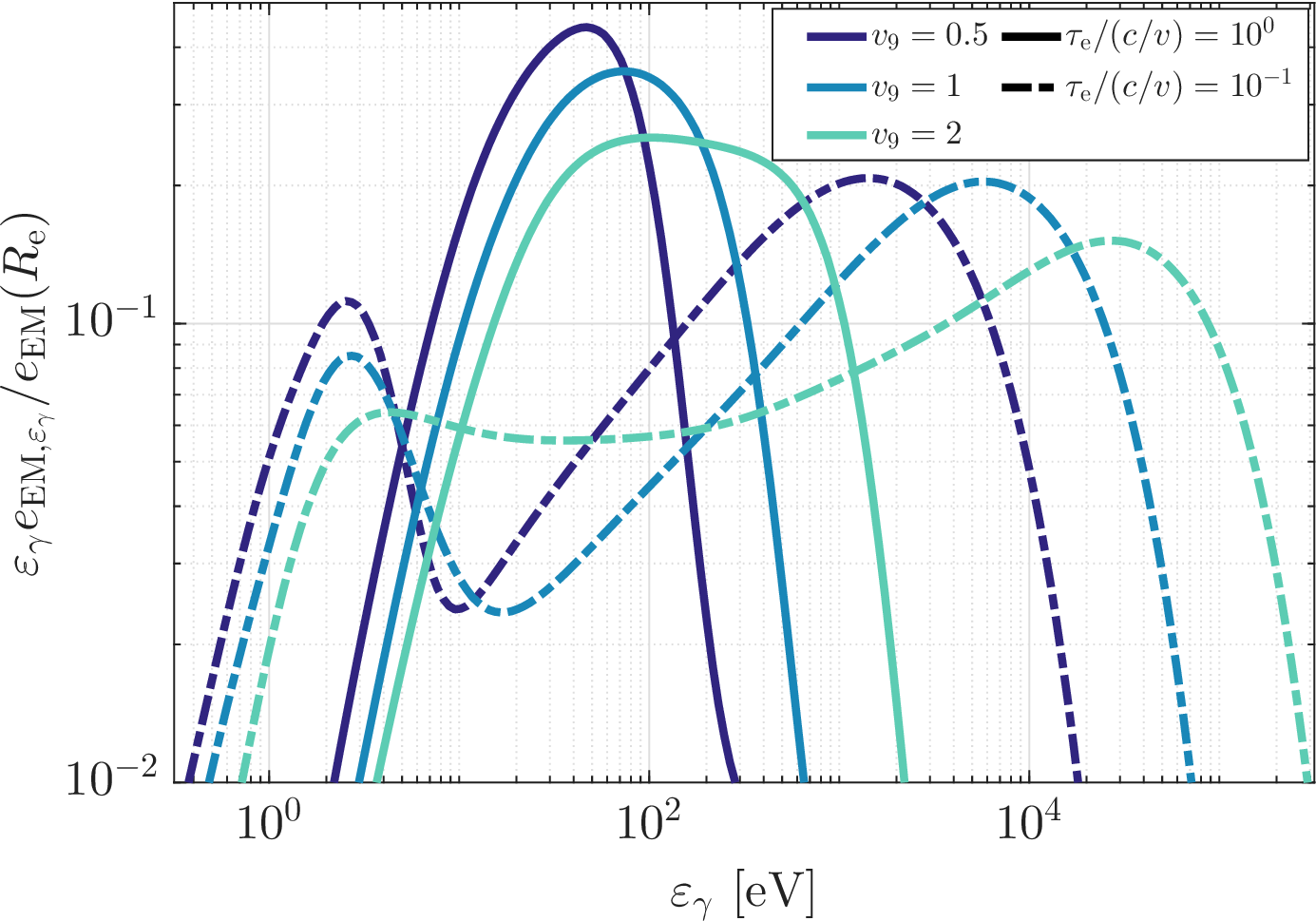}
    \caption{The EM energy distribution from \citet{wasserman_optical_2025}, at $r_{\rm sh}=R_{\rm e}$, normalized to the bolometric energy density, Equation (\ref{eq:e_EM}), for different $v_9$ and $\tau_{\rm e}/\left(c/v\right)=10^0,10^{-1}$. The curves are shown for $M_{\rm e}=10^{-1}\,M_\odot$, and the variations within an order of magnitude in $M_{\rm e}$ change the curves by an amount comparable to the line width.} 
    \label{fig:e_EM_ep}
\end{figure}

\begin{figure}
    \centering
    \includegraphics[width=8.5cm]{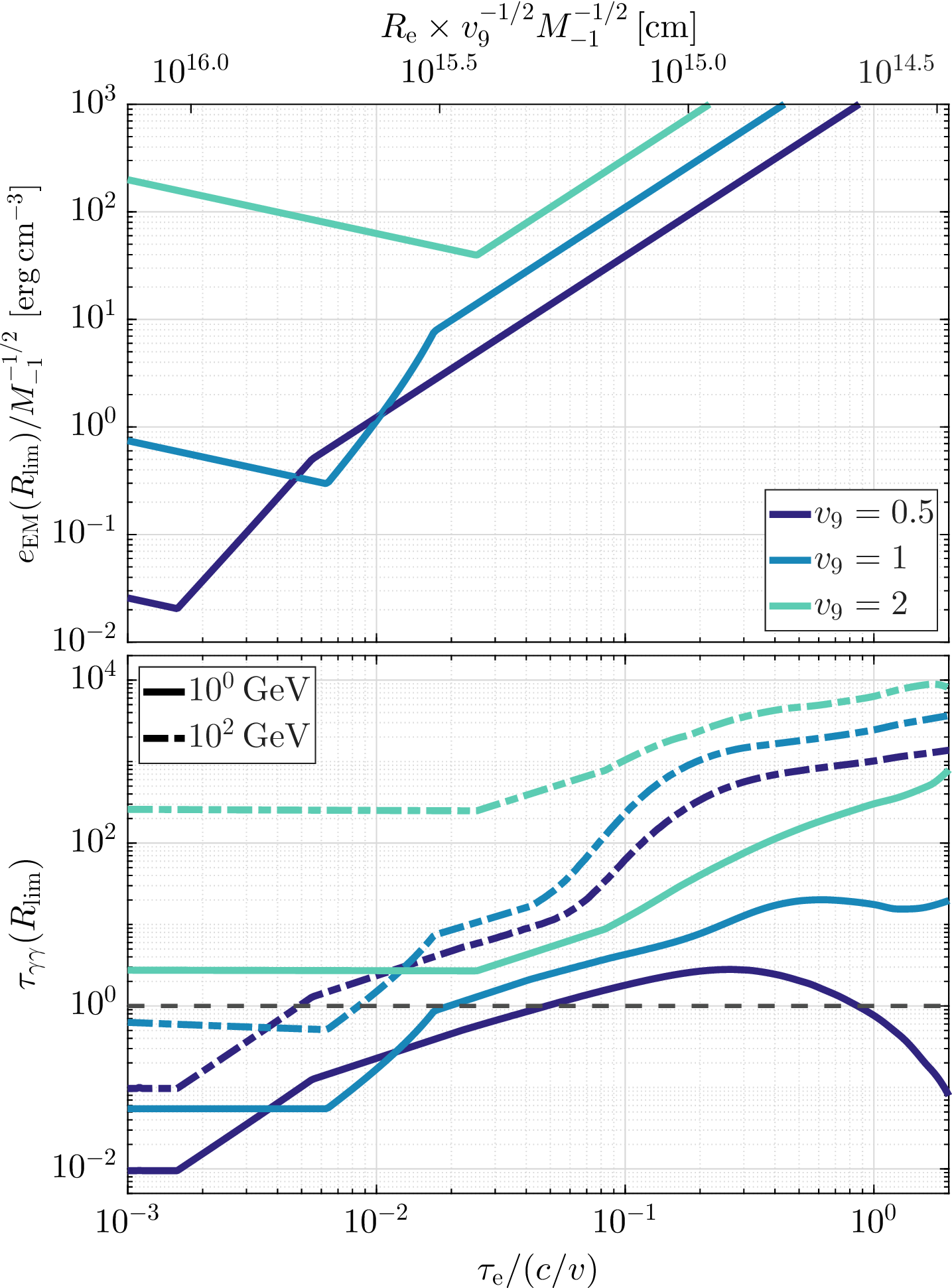}
    \caption{The upper panel shows the EM bolometric energy density, from Equation (\ref{eq:e_EM}), and the lower panel shows the $\gamma\gamma$ optical depth (for energies $10^0,10^2$ GeV), from Equation (\ref{eq:tau_gg}), both evaluated at $R_{\rm lim}$, as functions of the characteristic CSM optical depth, for different $v_9$ values. The upper $x$ axis shows the corresponding $R_{\rm e}$ values. The axes are normalized such that the EM energy density is exact for all $M_{\rm e}$, while the $\gamma\gamma$ optical depth is shown for $M_{\rm e}=10^{-1}\,M_\odot$, and the variations within an order of magnitude in $M_{\rm e}$ change the curves by an amount comparable to the line width. The horizontal dashed line indicates $\tau_{\gamma\gamma}=1$. We use $\kappa_{\rm T}=0.34\,{\rm cm^2\,g^{-1}}$, $\epsilon_{\rm CR}=10^{-1}$ and $\epsilon_B=10^{-2}$.}
    \label{fig:tau_gg}
\end{figure}

\begin{figure}
    \centering
    \includegraphics[width=8.5cm]{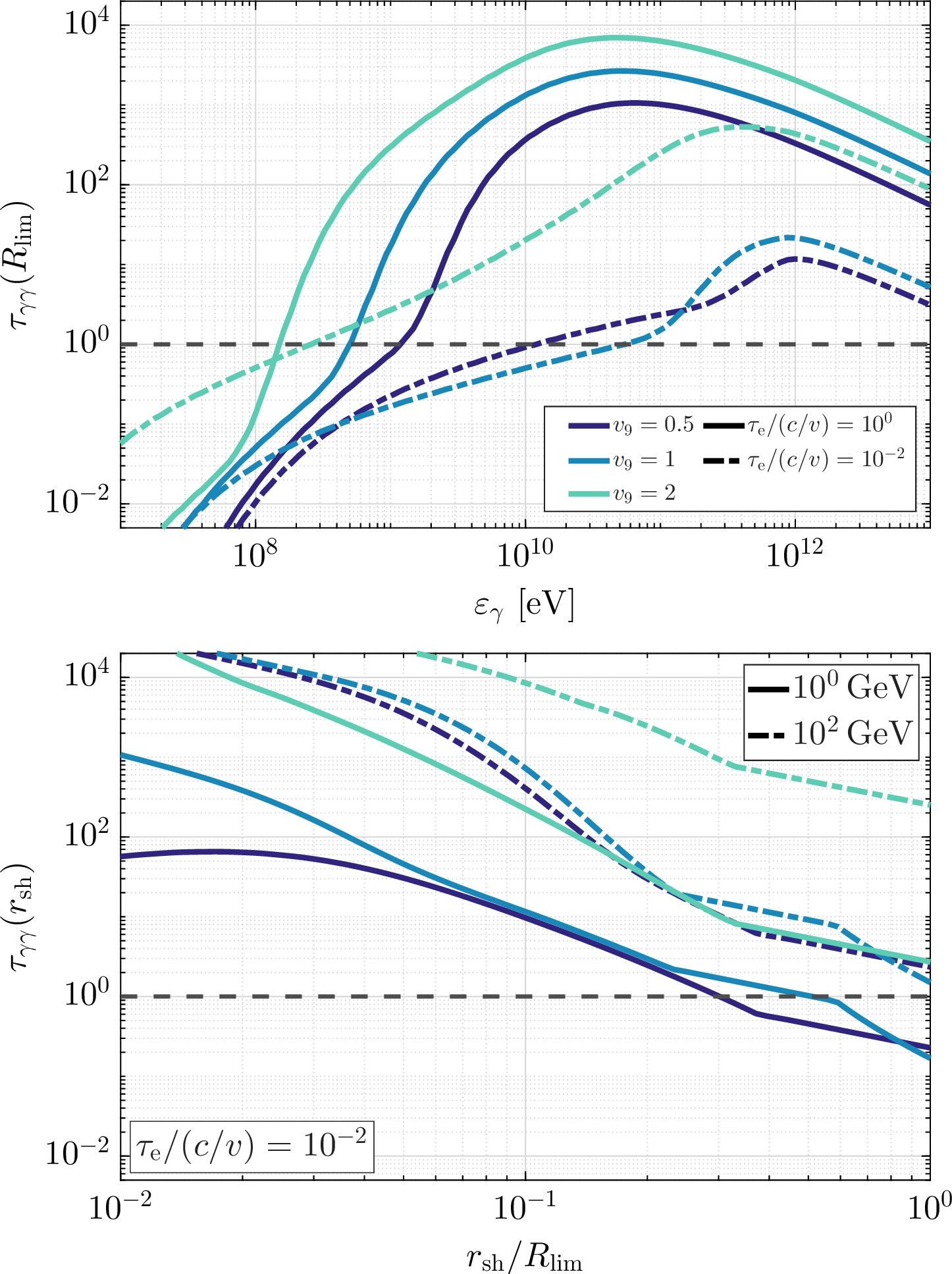}
    \caption{The $\gamma\gamma$ optical depth, in the upper panel evaluated at $R_{\rm lim}$ as a function of gamma-ray energy (for $\tau_{\rm e}/\left(c/v\right)=10^0,10^{-2}$) and in the lower panel as a function of $r_{\rm sh}$ (for $\varepsilon_{\gamma}=10^0,10^2$ GeV and $\tau_{\rm e}/\left(c/v\right)=10^{-2}$), for different $v_9$ values. $M_{\rm e}=10^{-1}\,M_\odot$ is shown, and variations within an order of magnitude in $M_{\rm e}$ change the curves by an amount comparable to the line width. The horizontal dashed line indicates $\tau_{\gamma\gamma}=1$. We use $\kappa_{\rm T}=0.34\,{\rm cm^2\,g^{-1}}$, $\epsilon_{\rm CR}=10^{-1}$ and $\epsilon_B=10^{-2}$.}
    \label{fig:tau_gg_ep}
\end{figure}

Figure \ref{fig:e_EM_ep} shows the numerical solution of the EM energy distribution from our earlier work \citep{wasserman_optical_2025}. Figure \ref{fig:tau_gg} shows the $\gamma\gamma$ pair-production optical depth, and the EM bolometric energy density, evaluated at $R_{\rm lim}$, as functions of the characteristic CSM optical depth. Figure \ref{fig:tau_gg_ep} shows the $\gamma\gamma$ optical depth as a function of gamma-ray energy and shock radius.

High-energy photons produced alongside neutrinos interact with the intense UV/X-ray photon field via the Breit-Wheeler pair-production process ($\gamma\gamma\to e^+e^-$). Since the radiation radial profile follows a power law, the optical depth is dominated by interactions at the smallest radius, so a local estimate of the optical depth is sufficient \citep[see, e.g.,][]{murase_high-energy_2019}. We approximate the cross section as having a collision angle $\theta=\pi$
\begin{equation}
\label{eq:tau_gg}
    \tau_{\gamma\gamma}(r_{\rm sh},\varepsilon_\gamma) = r_{\rm sh} \int d \varepsilon_\gamma'\frac{e_{\rm EM,\varepsilon_\gamma'}\left(r_{\rm sh}\right)}{\varepsilon_\gamma'}\sigma_{\gamma\gamma}(\varepsilon_\gamma,\varepsilon_\gamma'),
\end{equation}
where $e_{\rm EM,\varepsilon_\gamma}\left(r_{\rm sh}\right)$ is the background photon energy distribution (see Figure \ref{fig:e_EM_ep}) and $\sigma_{\gamma\gamma}$ is the $\gamma\gamma$ cross-section\footnote{The background photon emission is assumed to be isotropic, which will introduce an order unity correction. At larger radii, the propagating photon distribution becomes increasingly anisotropic, but this will introduce only a minor correction, as the optical depth is largely dictated by the emission radius.}, with a threshold energy of $\varepsilon_\gamma'>m_e^2c^4/\varepsilon_\gamma$ (approximately $3$($300$) eV for $100$($1$) GeV gamma-ray). Notice that the EM bolometric energy density at $R_{\rm lim}$ (presented in the upper panel of Figure \ref{fig:tau_gg}) decreases steeply with a decreasing characteristic CSM optical depth when $R_{\rm ad}<R_{\rm lim}$, and then increases when $R_{pp}<R_{\rm lim}$ (which happens earlier for $v_9>1.6$), see Section \ref{sec:formulation}. We evaluate the $\gamma\gamma$ optical depth at $R_{\rm lim}$ as most of the neutrino/gamma-ray emission occurs at this scale, and as at smaller radii the $\gamma\gamma$ optical depth is even larger.

For $\tau_{\rm e}/\left(c/v\right)\gtrsim10^{-2}$, we have $e_{\rm EM}\left(R_{\rm lim}\right)=\frac{1}{2}\rho_{\rm e}\frac{v^3}{c}$ (see Equation (\ref{eq:e_EM})), and if we crudely assume a monochromatic EM distribution of energy $\varepsilon'_\gamma$ close to the energy threshold, and a fixed $\sigma_{\gamma\gamma}=0.3\,\sigma_{\rm T}$, we get 
\begin{equation}
    \tau_{\gamma\gamma}\approx R_{\rm e}\frac{\frac{1}{2}\rho_{\rm e}\frac{v^3}{c}}{\varepsilon'_\gamma}0.3\,\sigma_{\rm T}=0.3\,\frac{\tau_{\rm e}}{c/v}\kappa_{0.34}^{-1}v_9^2\frac{m_ec^2}{{\varepsilon'_\gamma}}.
\end{equation}
If $\tau_{\rm e}/\left(c/v\right)$ is of order unity, we see that for $100$ GeV gamma rays interacting with $10$ eV photons that carry a sizable fraction of the energy (see Figure \ref{fig:e_EM_ep}), $\tau_{\gamma\gamma}\approx 10^4$. This is consistent with the results of the previous work~\citep[see Figure~5 of][]{murase_interacting_2024}.  
For $1$ GeV gamma rays interacting with $1$ keV photons that carry a sizable fraction of the energy, $\tau_{\gamma\gamma}\approx 10^2$.
 
The more detailed interplay between the changing EM energy density and its spectral shape (see upper panel of Figure \ref{fig:tau_gg} and Figure \ref{fig:e_EM_ep}) results in a typically opaque system for $\gamma\gamma$ pair-production all the way down to $1$ GeV gamma rays that interact with X-ray photons. For sufficiently extended CSM of $\tau_{\rm e}/\left(c/v\right)\lesssim10^{-2}-{10}^{-1}$, gamma rays can escape, as the radiation density decreases due to the transition of the shock from the radiative to the adiabatic regime (Equations (\ref{eq:e_EM})-(\ref{eq:R_ad})). For this reason, at high velocities ($v_9>1$), the system remains opaque even at a much smaller characteristic CSM optical depth, as the $pp$-inefficiency radius is smaller than the adiabatic radius. The $\tau_{\gamma\gamma}=1$ crossing is indicated also in Figure \ref{fig:flux_10TeV}. 

Thus, we expect that interacting SNe could be promising gamma-ray emitters if the CSM is sufficiently extended, as demonstrated in \cite{murase_interacting_2024}. Although we leave the determination of the luminosity and spectrum of the escaping gamma rays in this case to future work, our results clearly show the importance of detailed spectral energy distributions, especially in the X-ray range, for accurately assessing the detectability of gamma rays.

\section{Relation Between Neutrino and Electromagnetic Emission}\label{sec:nu_em_relation}

\begin{figure}
    \centering
    \includegraphics[width=8.5cm]{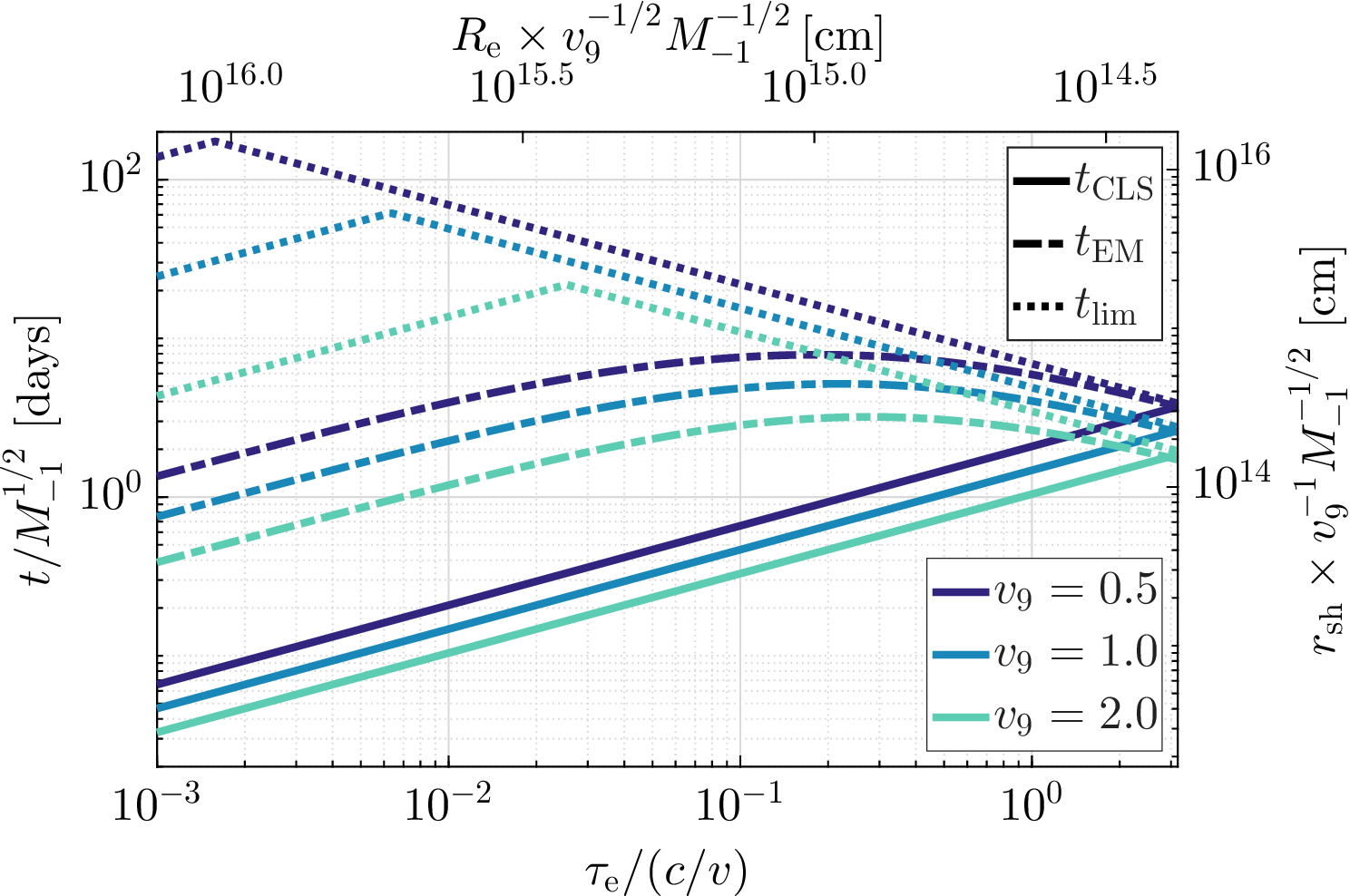}
    \caption{The CLS onset time $t_{\rm CLS}\equiv R_{\rm CLS}/v$, Equation (\ref{eq:R_CLS}), the EM light curve peak time, Equation (\ref{eq:t_EM}), and the end of the neutrino light curve time $t_{\rm lim}\equiv R_{\rm lim}/v$, Equation (\ref{eq:R_lim}), as a function of the characteristic CSM optical depth, for different $v_9$ values. The upper $x$ axis shows the corresponding $R_{\rm e}$ values, and the right $y$ axis the corresponding shock radius. The axes are normalized such that the curves are exact for all $M_{\rm e}$. We use $\kappa_{\rm T}=0.34\,{\rm cm^2\,g^{-1}}$, $\epsilon_{\rm CR}=10^{-1}$ and $\epsilon_B=10^{-2}$.}
    \label{fig:nu_timescales}
\end{figure}

\begin{figure}
    \centering
    \includegraphics[width=8.5cm]{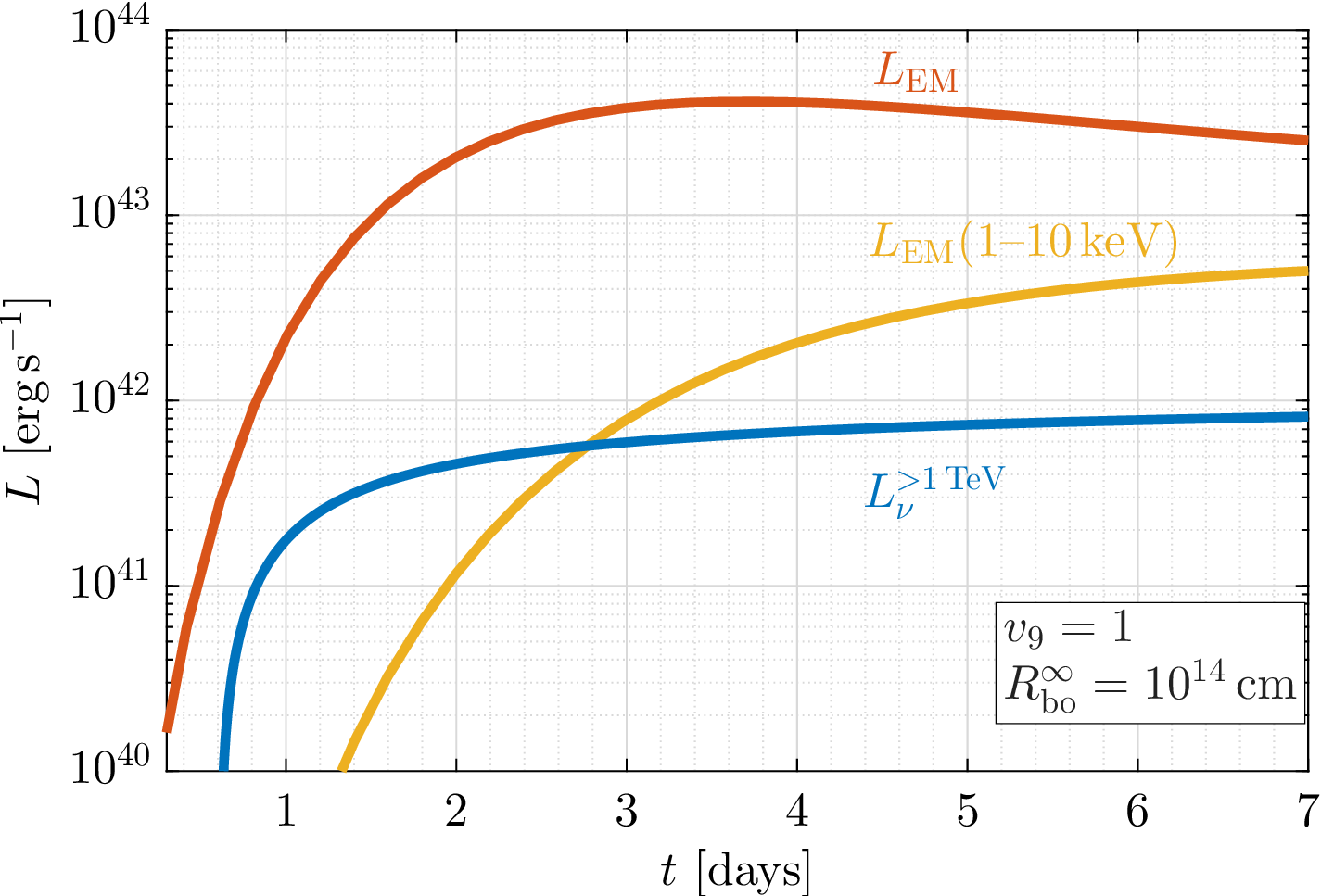}
    \caption{The $>1$ TeV neutrino and EM (bolometric and X-ray) light curves for an infinite CSM limit (see Section \ref{sec:formulation}). We use $\kappa_{\rm T}=0.34\,{\rm cm^2\,g^{-1}}$, $\epsilon_{\rm CR}=10^{-1}$ and $\epsilon_B=10^{-2}$.}
    \label{fig:EM_lightcurves}
\end{figure}

Figure \ref{fig:nu_timescales} shows different timescales characterizing the neutrino and EM light curves, as a function of the characteristic CSM optical depth, and Figure \ref{fig:EM_lightcurves} shows the neutrino and EM light curves for an infinite CSM limit (see Section \ref{sec:formulation}).

We see that for parameters of $\tau_{\rm e}/\left(c/v\right)\gtrsim 10^{-1}$, which are suggested by modeling of bolometric light curves, a significant fraction of the integrated energy of neutrinos is emitted prior to the peak of the EM light curve (as the photons are delayed by the diffusion timescale through the CSM). For stacking analyses, the relevant neutrino window should therefore begin before the EM peak and be tied to the inferred CSM parameters from the EM model (see Section \ref{sec:discussion}).

If the dense CSM is sufficiently extended ($\tau_{\rm e}/\left(c/v \right)\lesssim0.1-1$), the RMS-to-CLS transition pushes the photon characteristic energy from UV to X-ray on the breakout timescale of days, with a negligible suppression of the X-ray photons by the upstream plasma \citep{wasserman_optical_2025}. For more compact CSM distributions, the CLS phase may begin only beyond $R_{\rm e}$ of the dense CSM, and will produce a fainter X-ray signal by the shock propagation in a lower-density extended wind/CSM.

\section{Summary and Discussion}\label{sec:discussion}

In this paper, we analyzed the high-energy neutrino emission from nonrelativistic SN shock breakouts through compact CSM. We showed that both $pp$ and $p\gamma$ pion-production may be important for CR proton energy losses (Equations (\ref{eq:t_pp})-(\ref{eq:t_pg}) and Figure \ref{fig:p_timescales}) limiting the maximal proton energy to $\varepsilon_p^{\max}\lesssim20\,{\rm PeV}\,\epsilon_{B,-2}^{1/2}$, and the corresponding maximal neutrino energy to $\varepsilon_\nu^{\max}\lesssim1\,{\rm PeV}\,\epsilon_{B,-2}^{1/2}$ (Equation (\ref{eq:ep_max}) and Figure \ref{fig:E_nu_tot}). We showed that below the $pp$-inefficiency radius (Equation (\ref{eq:R_pp})), where all of the CR proton energy is efficiently converted into pions, about half of it is converted to neutrinos (Equation (\ref{eq:L_nu})), as the pions and muons decay timescales are sufficiently short (Figure \ref{fig:mu_time}). Therefore, the total neutrino production efficiency (defined as the part of the interaction energy that is converted to neutrinos, $E_\nu^{\rm tot}/E_{\rm sh}$) can be as high as $5\%\,\epsilon_{\rm CR,-1}$, and is suppressed for both too compact CSM configurations, $\tau_{\rm e}/\left(c/v\right)\rightarrow3.3$, where the CLS onset radius is close to the CSM radial extent, and for highly extended CSM configurations, $\tau_{\rm e}/\left(c/v\right)\ll 1$, where an increasing fraction of the CSM mass is at a density too low for an efficient $pp$ pion-production, see the suppression function $f\left(r\right)$ in Equations (\ref{eq:E_nu})-(\ref{eq:S(r)}) and Figure \ref{fig:E_nu_tot}. 

For CSM compactness of $\tau_{\rm e}/\left(c/v\right)\gtrsim10^{-1}$, a significant fraction of the neutrino emission is released prior to the peak of the EM light curve (Figures \ref{fig:L_nu_tot}, \ref{fig:nu_timescales} and \ref{fig:EM_lightcurves}), as the photons are delayed by the diffusion timescale through the CSM (Equations (\ref{eq:t_diff})-(\ref{eq:t_EM})). We derived a simple and accurate universal description of the neutrino energy distribution accumulated as the shock traverses the CSM, $\varepsilon_\nu^2dn_\nu/d\varepsilon_\nu=E_\nu^{\rm tot}\left(1-\varepsilon_\nu/\varepsilon_\nu^{\max}\right)/\ln\Lambda$ (Equations (\ref{eq:spectrum_approx1})-(\ref{eq:spectrum_approx2}) and Figure \ref{fig:nu_spect}).

The $10$ TeV neutrino flux obtained at Earth, for a local rate of $10^{-4}\,{\rm Mpc^{-3}\,yr^{-1}}$ identical sources of SN breakouts through compact CSM, may constitute a sizable fraction of the observed background for $10^{-2}\lesssim\tau_{\rm e}/\left(c/v\right)\lesssim 0.3$, where the neutrino efficiency does not suffer from the suppression mentioned above (Equation (\ref{eq:flux_approx}) and Figure \ref{fig:flux_10TeV}). The interplay between the evolving radiation energy density and its spectral shape (see upper panel of Figure \ref{fig:tau_gg} and Figure \ref{fig:e_EM_ep}) results in a typically opaque system for $\gamma\gamma$ pair-production all the way down to $1$ GeV gamma rays that interact with X-ray photons. For sufficiently extended CSM configurations, $\tau_{\rm e}/\left(c/v\right)\lesssim10^{-2}-10^{-1}$, gamma rays can escape, as the radiation density decreases due to the transition of the shock from the radiative to the adiabatic regime, which we explored in this work, see Equations (\ref{eq:e_EM})-(\ref{eq:R_ad}).

High-energy neutrino detectors of $\sim1\,{\rm km^2}$ area should be able to detect, on average, $>1$ muon-neutrinos only for very nearby events \citep{murase_new_2018,kheirandish_detecting_2023}. Assuming 1\% efficiency of producing $1-100$ TeV neutrinos, for which the effective detector area is $\approx10^{-6}\left(\varepsilon_\nu/1\,{\rm TeV}\right)A$, where $A=10^{10}\,A_{10}\,{\rm cm^2}$ is the geometrical cross-section of the detector, the average number of muon-neutrino detections from an event at distance $d$ is \citep[e.g.,][]{katz_x-rays_2011}
\begin{equation}
\begin{split}
    N_\mu&=\frac{10^{-6}\,A}{4\pi\,d^2}\frac{E_{\nu,1-100\,{\rm TeV}}/3}{1\,{\rm TeV}}\\
    &\sim 1\,A_{10}\left(\frac{d}{5\,{\rm Mpc}}\right)^{-2}\left(\frac{f_{1-100\,{\rm TeV}}}{1\%}\right)M_{-1}v_9^2.
\end{split}
\end{equation}
Type II SNe are expected to occur within $5$ Mpc at a rate of $\sim1/20$ yr. For $\sim10\,{\rm km^2}$ detectors (which are under planning and construction, IceCube-Gen2, \cite{grant_neutrino_2019}; P-ONE, \cite{agostini_pacific_2020}; TRIDENT, \cite{ye_multi-cubic-kilometre_2023}), the horizon of a single detection on average will expand to $15$ Mpc, where the rate of SNe increases to $\sim 1\,{\rm yr^{-1}}$. An association of a single 10 TeV muon-neutrino with a SN at the horizon distance within a week of the explosion would be significant at $\approx99.9\%$ confidence level: for an atmospheric neutrino background rate of $\approx2 \times 10^3\,{\rm yr^{-1}}$ \citep{icecube_collaboration_evidence_2022}, the expected number of atmospheric-background events in a 7-day, 1 ${\rm deg^2}$ window is $\approx10^{-3}$. 

For SN 2023ixf and SN 2024ggi at $d\sim7$ Mpc, whose early light curves are thought to be powered by a $v=10^9\,{\rm cm\,s^{-1}}$ shock breakout through dense CSM of $M_{\rm e}\sim0.03\,M_\odot$ \citep[e.g.,][]{hiramatsu_discovery_2023,zimmerman_complex_2024,jacobson-galan_sn_2024}, we expect $N_\mu\sim0.1$, consistent with the nondetections. In fact, if the CSM is confined to $R_{\rm e}\sim10^{14.5}$ cm as some of the interpretations suggest, then $\tau_{\rm e}/\left(c/v\right)\sim0.5$, which suppresses the efficiency by a factor $\sim3$ (see Figure \ref{fig:E_nu_tot}). The high-energy, $>1$ GeV, gamma rays from the pions decay should have approximately two-thirds of the neutrino luminosity, $\approx10^{42}\,{\rm erg\,s^{-1}}$, which is about twice the Fermi upper limit in the first days of SN 2023ixf \citep{marti-devesa_early-time_2024}. This may be consistent with the large $\gamma\gamma$ pair-production optical depth calculated in our analysis for such compact CSM distributions.

Even when considering a steeper proton index of $s=2.2$ (see Section \ref{sec:nu_emission}), normal type II SN explosions may contribute $\sim 10\%\left(M_{\rm e}/0.03\,M_\odot\right)$ of the observed background if such large CSM masses are common, and type IIn SNe may produce another $\sim 10\%(M_{\rm e}/0.3\,M_\odot)$ of the background, for $\epsilon_{\rm CR}=0.1$. The distribution of CSM parameters in the SN population is unknown and does not yet allow for a more precise estimation.

As the capabilities of rapid transient searches improve, both from the ground (e.g., GOTO, \cite{steeghs_gravitational-wave_2022}, and LAST, \cite{ofek_large_2023}) and from space with the expected launch of the wide-field UV space telescope ULTRASAT \citep{shvartzvald_ultrasat_2024}, a systematic detection of many SNe of all types at early, $<1$\,d time will be possible. The systematic early observations of SNe will enable the use of the results derived in this paper to determine the CSM properties and, hence, the neutrino flux and spectrum expected from CSM breakouts, and their contribution to the neutrino background.

\begin{acknowledgments}
The research of TW and EW is partially supported by ISF, Minerva and Segre grants. The work of KM is supported by the NSF Grants No.~AST-2108466, No.~AST-2108467, and No.~AST-2308021. 
\end{acknowledgments}

\bibliography{references}

\end{document}